\documentclass[aps,
 twocolumn,
superscriptaddress,notitlepage]{revtex4-1}
\usepackage{graphicx}
\usepackage{bm}
\usepackage{amsmath}
\usepackage{amsfonts}
\usepackage{amssymb}
\usepackage{array}
\usepackage{xcolor}
\usepackage{dcolumn}
\usepackage{longtable}
\usepackage{hyperref}
\usepackage{natbib}
\usepackage{stmaryrd}%
\usepackage{color}
\usepackage{wasysym} % for hexagon and star of david in equation
\usepackage{siunitx}
\usepackage{enumitem}
\usepackage{lipsum}
\usepackage{amsmath,amsthm}
\usepackage{amsfonts}
\usepackage{amssymb}
\usepackage{bbold}
\usepackage{amssymb}
\usepackage{epstopdf}
\usepackage{times}
\usepackage{mathrsfs}
\usepackage{color}
\usepackage{gensymb}
\usepackage{times}
\usepackage{subfigure}
\usepackage[]{graphicx}
\usepackage{amssymb}
\usepackage{amsmath}
\usepackage{bm}
\usepackage{bm,amsmath}
\usepackage[normalem]{ulem}

\definecolor{mygreen}{RGB}{98,174,37}
\definecolor{myred}{RGB}{211,0,45}
\definecolor{myblue}{RGB}{0,126,148}

\def\phi{\varphi}

\def\incep{\left\{\begin{array}{ll} }
 \def\termin{\end{array}\right. }

\def\la2{\lambda^2}

\begin{document}
%\title{An accurate computation of true stress in compression experiments in the presence of localized deformation}

 \title{Self-induced  marginality in plastically deformed crystals}
%\title{Two-stage yielding   of pristine crystals}
%\title{Quasi-amorphous crystal}
%\title{\textcolor{blue}{Strain induced transition from perfect crystal to marginally stable solid}}
\author{O.U. Salman}
\affiliation{
LSPM, CNRS UPR3407, Université Sorbonne Paris Nord, 93400, Villateneuse, France}
\affiliation{
Lund University, Department of Mechanical Engineering Sciences, SE-221 00 Lund, Sweden}
\author{A. Ahadi}
\affiliation{
Lund University, Department of Mechanical Engineering Sciences, SE-221 00 Lund, Sweden}
\author{ L. Truskinovsky}
\affiliation{
PMMH, CNRS UMR 7636 ESPCI PSL, 10 Rue Vauquelin, 75005, Paris, France}

\date{\today}
\begin{abstract}
Quasi-brittle plastic yielding is a salient feature  of well-annealed glassy   materials. Here we show that  the same behavior is characteristic of   perfect crystals after they experience mechanically  driven  elastic instability leading to massive nucleation of dislocations.   We argue that such 'preparation' effectively converts  an  atomic configuration from  crystalline  to quasi-amorphous.   To understand the nature of  the subsequent intermittent mechanical response    we  study   a  model 2D crystal   subjected to AQS  driving and show that   both pre- and post-yield  dislocation avalanches  exhibit power law statistics with  similar  exponents  indicative of   self-induced marginal stability. 
\end{abstract}
\maketitle
\noindent

The complexity of  plastic response of  strain-driven crystalline  solids is revealed  by   intermittent stress fluctuations accompanied by scale-free  spatial organization of dislocations ~\cite{
%Zaiser2006-gk,
%jian2024prediction,ortiz1999plastic,
Papanikolaou2017-ld,
%kosterlitz2016kosterlitz,khantha1997mechanism,
hochrainer2014continuum,
%langer2019statistical,
Derlet2016-mp,ispanovity2010submicron}. The same general  features  are also typical   of amorphous plasticity  ~\cite{
%kamani2024brittle,
berthier2024yielding,Shang2020,
%shang2024yielding,
singh2020brittle,
%richard2021brittle,pollard2022yielding,
 Lin2014-bx}. 
 %parisi2017shear,
%derlet2021micro
To build a  conceptual link between the two phenomena,  we  study  in this Letter a relatively transparent prototypical example of a  2D square crystal  subjected to athermal quasistatic (AQS) proportional loading by shear strain applied  along one of the  slip directions.  To prepare a 'generic' crystal we start from a  pristine dislocation-free   crystal and shear it  it to the point of mechanical instability which is resolved through   massive dislocation nucleation.  We argue that such 'preparation' effectively converts  an  atomic configuration from  crystalline  to quasi-amorphous, e.g.    \cite{
%Lehtinen2016-qy,Sethna2017,
Ovaska2017}. This term is used to  refer  to a solid  that is structurally crystalline but mechanically amorphous in the sense that its  mechanical response is dominated by non-affine relaxations, e.g.  \cite{Zaccone2011-er}.

We observe that  plastic flow  of   such quasi-amorphous crystals begins with a   hardening stage  involving    small spatially localized dislocation rearrangements.  This   'microplasticity' regime   is terminated    abruptly with a   quasi-brittle     yield     
  leading to  the   formation of  multi-grain pattern  dominated by  global  shear bands. The post-yield plastic flow is characterized by  a stress plateau with superimposed fluctuations representing broadly distributed  dislocation  avalanches.  
Similar   quasi-brittle  yield has been observed  in well-annealed glasses  \cite{
%Nicolas2018-iy,Rodney2011-ld,popovic2018elastoplastic,
Tanguy2021-ap,
%TALAMALI2012275, 
Parley2024-jb,
%karmakar2010statistical,
ozawa2018random}.  
%Divoux2024-ks
 Despite the absence of dislocations in amorphous plasticity,     both phenomena can be viewed as representing the  evolution of elastic incompatibility   with  underlying elementary mechanical events   represented by  shear eigenstrains \cite{Picard2004,Budrikis2017-ex,
%Cai2006-fe, Moshe2015-rm,
Baggioli2022-nx}.
%Bowick2009-bl,
%Katanaev1992-kl,Sozio2020-tg

%To reveal the universal features of the  criticality exhibited by crystal and amorphous plasticity, we perform a detailed study of  dislocation dynamics  in  a prototypical  athermal square crystal  subjected to   quasistatic loading by shear strain applied  along one of the  slip directions. 

Our main tool is a  novel mesoscopic tensorial model  (MTM) of crystal plasticity representing a conceptual trade-off between continuum  and atomic descriptions~\cite{Salman2011,Salman2012b,Baggio2019,Salman2021,perchikov2024,Baggio2023,Baggio2023-qu,Zhang2020-ax}. 
Crucially, it  allows one  to  capture the  intermittent nature of   dislocation avalanches while   resolving  topological (connectivity) restructuring  involved in    nucleation and annihilation of dislocations  without any specialized   phenomenological assumptions \cite{Weiss2021-tt}.
  In contrast to other  computational approaches to micro-scale plasticity, e.g. \cite{
  %Arsenlis2007-oz,
  Zepeda-Ruiz2021-bm,
  %Ponga2020-kd,Lafourcade2025-pc,
  Chan2010-qt,Salvalaglio2020-aw,
  %Zhong2025-df,
  Javanbakht2016-dr,Beyerlein2016-av,Koslowski2002-dn,Wang:2003zh,Devincre2008-mw,Ispanovity2014-ra,Gomez-Garcia2006-zn,Po2014-qu,
 % Lu2022-ai,Bertin2024-jw,Nasir-Tak2025-kq,
  Starkey2020-yv,Zhang2025-cm},
%  Acharya2022-ur,Acharya2006-du,Rodney1999-em,Van-Koten2011-tt,
%  Miller2002-pa,Zhang2018-in
  the MTM  allows one to deal with statistically meaningful number of dislocations while accounting  in a geometrically exact way for finite deformations ~\cite{Engel1986-tm,
 % Michel2001,schwarzenberger1972classification,
  pitteri2002continuum}.

 The main idea of the MTM approach is to  associate with material points   an effective energy landscape whose tensorial periodicity accounts for crystallographically-invariant deformations. The resulting Landau-type model  is  characterized by infinitely many equivalent energy wells \cite{Ericksen1970,
% ericksen1973loading,Ericksen1977,Ericksen1980,ericksen1989weak,
% parry1976elasticity,Parry1977,
 Parry1998,Folkins1991,Conti2004-sv,Bhattacharya2004,Salman2011,
 %Salman2012b,
 Baggio2019,
 %Salman2021,
 perchikov2024}
 %,Baggio2023,Baggio2023-qu,Zhang2020-ax
and   plastically deformed crystals emerge as a multi-phase mixtures with  dislocations playing the role of effective domain boundaries. The small scale regularization is  achieved by mesoscopic discretization. Despite operating with engineering concepts of stress and strain the resulting  computational code  correctly captures not only long-range elastic interactions of  dislocations but also  describes adequately their short-range interactions involved in   the formation of  complex dislocational entanglements, see also variations on the same theme in
% Elements of the MTM  approach can be   traced to 
 \cite{Denoual2010,Gao2019,Biscari2015-ci,Arbib2023}.
 %Gao2020,Gao2020a,Gao2020f,Gao2020c
 %see also  a parallel  recent development  in \cite{Arbib2023}. 
% Arbib2020,,Arbib2025
We also note that the  2D version of the   MTM,   used in our numerical experiments,  has some formal similarities with  a single-slip-plane   phase field  model of Peierls-Nabarro type, which was found fully adequate to represent  the critical nature of   plastic flows, e.g. \cite{Koslowski2002-dn, Koslowski2004-sa}.  The main difference of our approach is the use of  finite strain  bulk elasticity  compatible with  multi-slip  and capturing adequately lattice rotations which produced different values of critical exponents revealing the marginal stability of the underlying configurations.

%\textcolor{blue}{[Referee A/B: Technical Definitions] In our simple shear loading $\mathbf{F} = \mathbf{I} + \gamma \mathbf{e}_1 \otimes \mathbf{e}_2$, the stress component $\sigma_{12}$ is the conjugate stress to the primary loading mode. While $\sigma_{12}$ is coordinate-dependent, our MTM framework derives all quantities from the objective metric tensor $\mathbf{C}$, ensuring physical consistency under finite rotations. Furthermore, by "topological transition," we refer to discrete jumps of mesoscopic elements between equivalent $GL(2, \mathbb{Z})$-related energy wells, corresponding to quantized changes in the Burgers vector.}

\begin{figure}[h!]
\centering
\begin{tabular}{cc}
\includegraphics[height=3.6cm]{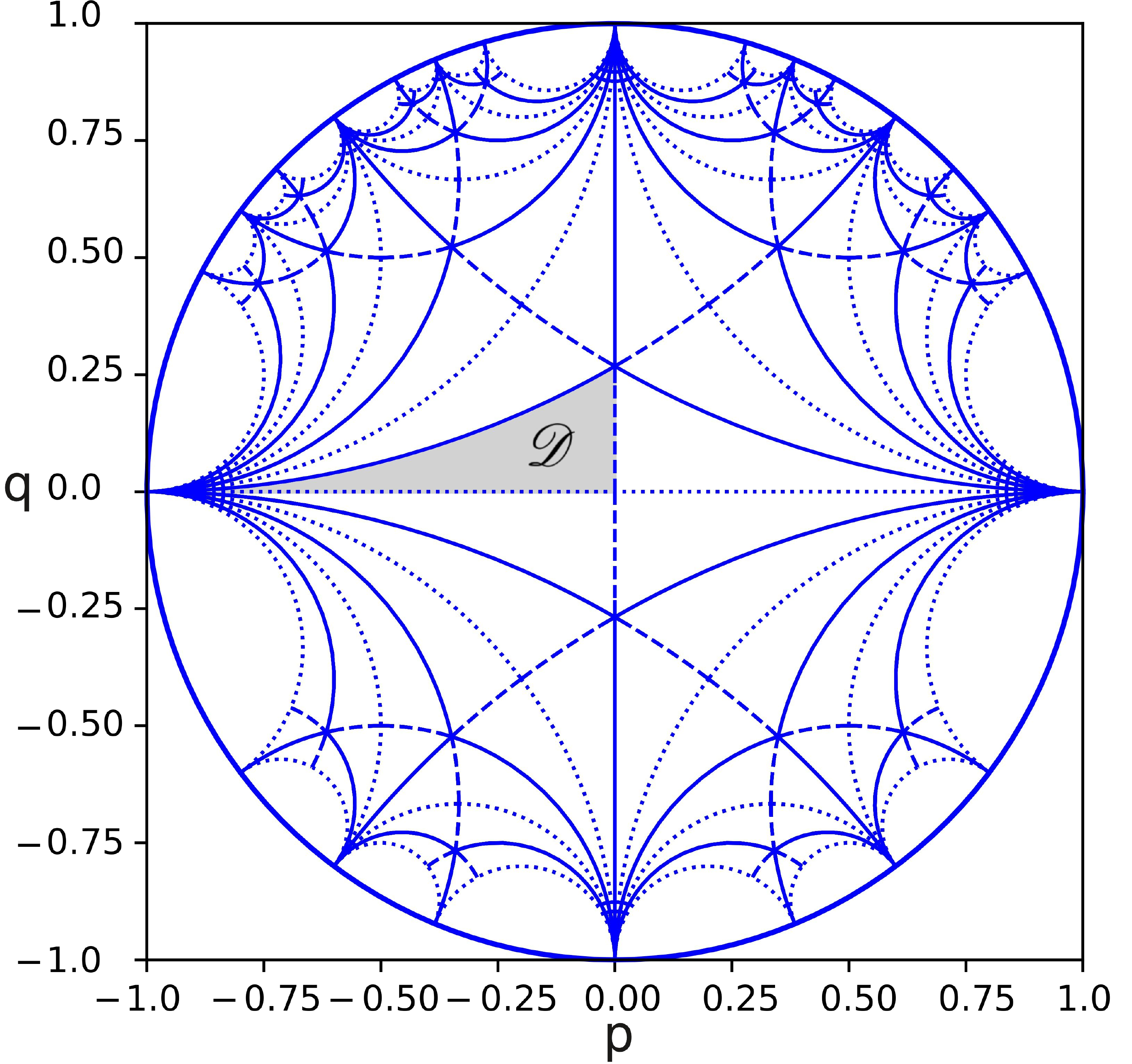} &
\raisebox{0.8mm}{\includegraphics[height=3.6cm]{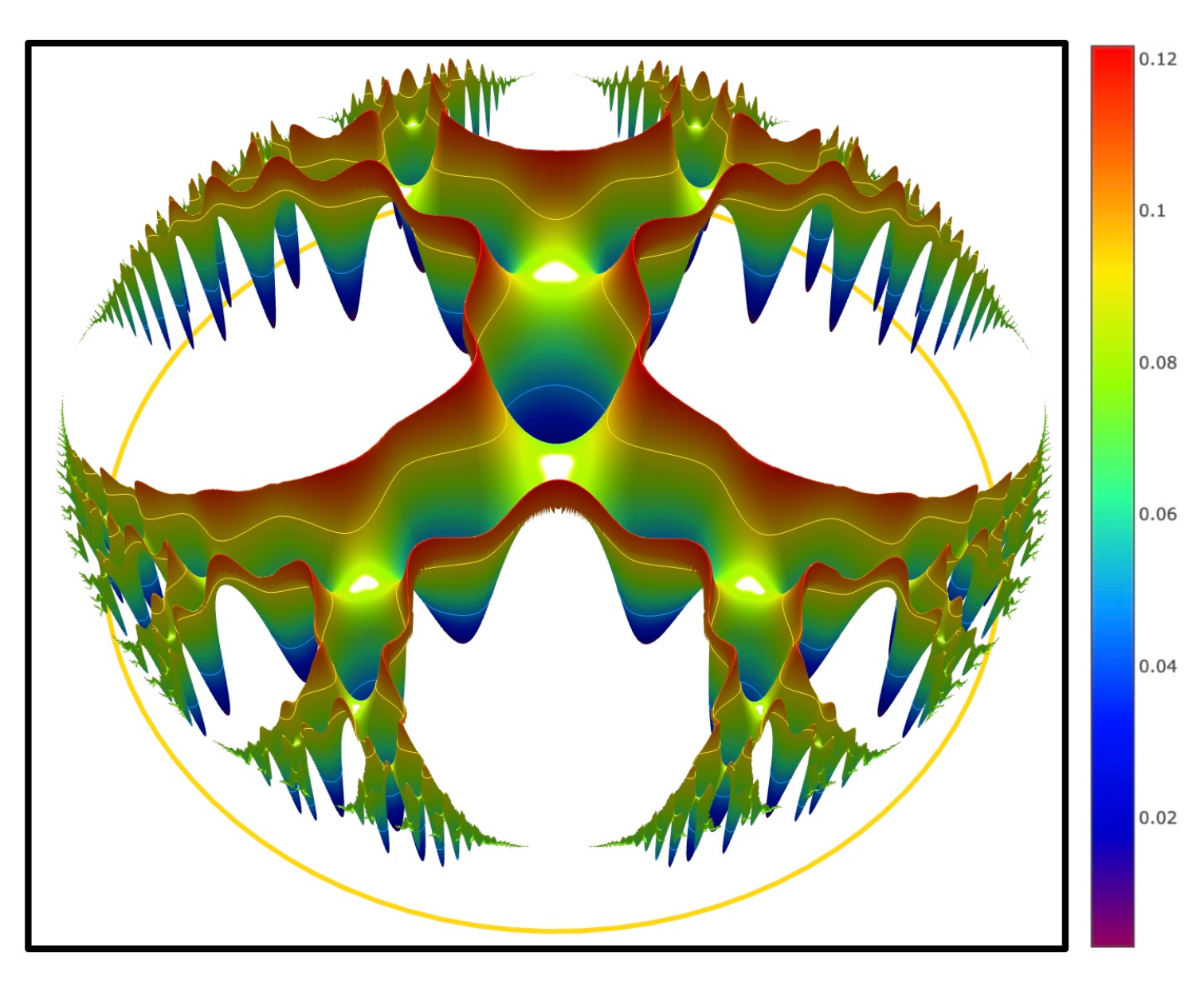}} \\
(a) & (b)
\end{tabular}
\caption{(a) $GL(2, \mathbb{Z})$-induced tessellation of the configuration space of metric tensors $C_{11}, C_{22},C_{12}$  with $\det \mathbf{C}=1$ stereographically projected on the  Poincar\'e disk  $p^2 + q^2 < 1$ where 
$p = t(C_{11} - C_{22})/2$, 
$q = tC_{12}$, 
and  $t = 2/(2 + C_{11} + C_{22})$.
The highlighted domain $\mathscr{D}$  is the fundamental (minimal periodicity) domain. (b) Representative energy landscape  with colors indicating the energy level.  Only the lowest energy levels are visible with the bottoms of the  wells corresponding to equivalent replicas of the reference square lattice.}
\label{fig:atlas0}
\end{figure}  

 Consider  the continuum deformation  field
%\begin{equation}
$\bf y = \bf y( \bf x),$
%\end{equation}
  where   $\bf y$ and $\bf x$  are  positions of material  points  in the current and reference configurations, respectively.
  Due to frame indifference  constraint, the strain energy density  $\phi$
  % $\phi$
  %= \phi(\bf C)$ 
 can  depend on the deformation gradient
  %\begin{equation}
 $\bf F = \nabla \bf y$
 %\end{equation}
only through the metric tensor  
%\begin{equation}
$ {\bf C} = {\bf F}^{T}{\bf F} $
which then  plays the role of a tensorial order parameter.  
 %\end{equation}
%  The rotational invariance of the strain energy density $\phi(\bf C)$ for ${\bf F} \in {\bf SO}(2)$ is ensured by the structure of the Cauchy-Green tensor $\bf C$. 
In MTM we assume that the function  $\phi(\bf C)$ additionally  respects the  symmetry of an underlying  Bravais lattice.  Such symmetry is assumed to be global extending beyond the conventional  point group and capturing lattice invariant shears. In  2D  it reduces to the constraint that 
%\begin{equation}
$ \phi({\bf C}) = \phi({\bf m}^T {\bf C} {\bf m}),$
% \end{equation}
   where   unimodular integer valued matrices   $\bf m$ belong  to  $ GL(2,\mathbb{Z})$ \cite{SM}. 
In view  of  this   symmetry, the surface $\det{\bf C}=1$  in the 3D space $(C_{11}, C_{22}, C_{12})$   is tessellated   into an infinite number of equivalent periodicity domains; if the  energy density  $ \phi({\bf C}) $ is known in one of  such  domains, it can be extended to the whole configurational space of tensors $\bf C$ by the  symmetry.  To  visualize the implied  tessellation, it is convenient to project stereographically the infinite surface $\det{\bf C}=1$  on a 
%(Poincar\'e) 
unit (Poincar\'e) disk,  see   Fig.~\ref{fig:atlas0}(a).
% \textcolor{red}{where  the equivalent square lattices are indicated by small black squares. WE REMOVED THE SQUARE DOTS}
% while red dots denote equivalent triangular (hexagonal) lattices. 
%where we show in gray  the fundamental (minimal  periodicity) domain $\mathscr D$. 
 %which is known in the mathematical literature on modular forms as a 'fundamental domain', 
  % is discussed in \cite{SM}.  
 In our numerical experiments we used  the  simplest piece-wise  polynomial  function  $\phi(\bf C)$ proposed in \cite{Conti2004-sv},
  % which is the simplest to ensure  stress continuity, 
   see  \cite{SM} for details. In 
   Fig.~\ref{fig:atlas0}(b)
  % Fig. \ref{fig:atlas1}  
   we illustrate the corresponding  multi-well  energy landscape. Here  the   equivalent energy  wells, which are somewhat distorted due to strereographic projection,  represent the equivalent  replicas of the reference square  lattice.  A characteristic feature of  such  a landscape is a network of low-energy valleys  corresponding  in our case to simple shears parallel to crystallographic slip planes. These valleys  represent  conventional plastic 'mechanisms' and are assumed to be flat in the classical continuum crystal plasticity.
   
%    Along such  valleys  the   energy wells  designate  equivalent replicas of the reference square  lattice. 

Since  the MTM preserves the energy barriers separating individual  wells along the low-energy valleys,   the function $\phi(\mathbf{C})$ ends up being highly nonconvex.
%, see Fig. \ref{fig:atlas1}. 
The corresponding  continuum scale-free Landau model is  then degenerate and needs to be regularized through the introduction of a cut-off length scale. This is   achieved through  the projection of the continuum problem on  a uniform mesoscale grid while associating the  nonconvex  elastic  energy   to  finite discrete   elements.   In other words,  the infinite dimensional space of continuum deformation fields is reduced to a finite dimensional set of compatible, piece-wise affine mappings,  with the  linear scale of an individual finite element  becoming a  physical parameter  \cite{SM}. An incremental elastic energy minimization in a system of this type subjected to  quasistatic loading will generically lead to a rich repertoire of instabilities with jump discontinuities representing  elastic branch switching events. The energy losses associated with such jumps contribute to  plastic dissipation; in the continuum limit the jumps   will merge  giving rise to  rate-independent plasticity \cite{Puglisi2005-lg}.

%The  driven system of this type will experience a rich repertoire of instabilities.   Specificaly,  An incremental elastic energy minimization of a system subjected to  quasistatic loading will generically lead to repeated jump discontinuities representing  branch switching events. The energy losses associated with such jumps will represent  plastic dissipation; in the continuum limit the jumps   will merge  giving rise to  rate-independent plasticity \cite{Puglisi2005-lg}.

 In our numerical experiments a square crystal  was represented by  $N\times N$ nodes,  with $N=100, 200, 400$. The nodes were arranged into  triangular finite elements.  We loaded the  system quasi-statically by  applying the affine displacement field 
 %\begin{equation}
 $ {\bf u}(\alpha,{\bf x})= (\bar{\bf F}(\alpha )-\mathbb{1}){\bf x} , $
  %\end{equation}
 with 
 $\bar{\bf F}(\alpha)={\bf I}+\alpha {\bf e}_1 \otimes  {\bf e}_2,$ 
%${\bf F}(\alpha,\theta)={\bf I}+\alpha {\bf R}(\theta){\bf e}_1 \otimes{\bf e}_2$,
 where $ {\bf e}_ i$, $i=1,2$,  is the orthonormal basis of the reference square lattice.  In this way we imposed  simple shear along one of the principal slip directions with shear amplitude $\alpha$ serving as the loading parameter. The loading was advance   in strain increments of order $10^{-6}$ and after each   increment the displacement field was updated by the energy minimization algorithm. Details of the discretization, of the  incremental  energy minimization algorithm and of  the implementation of the loading protocol under periodic  boundary conditions can be found  in \cite{SM}.

 \begin{figure}[h!]
\centering
\includegraphics[scale=.043]{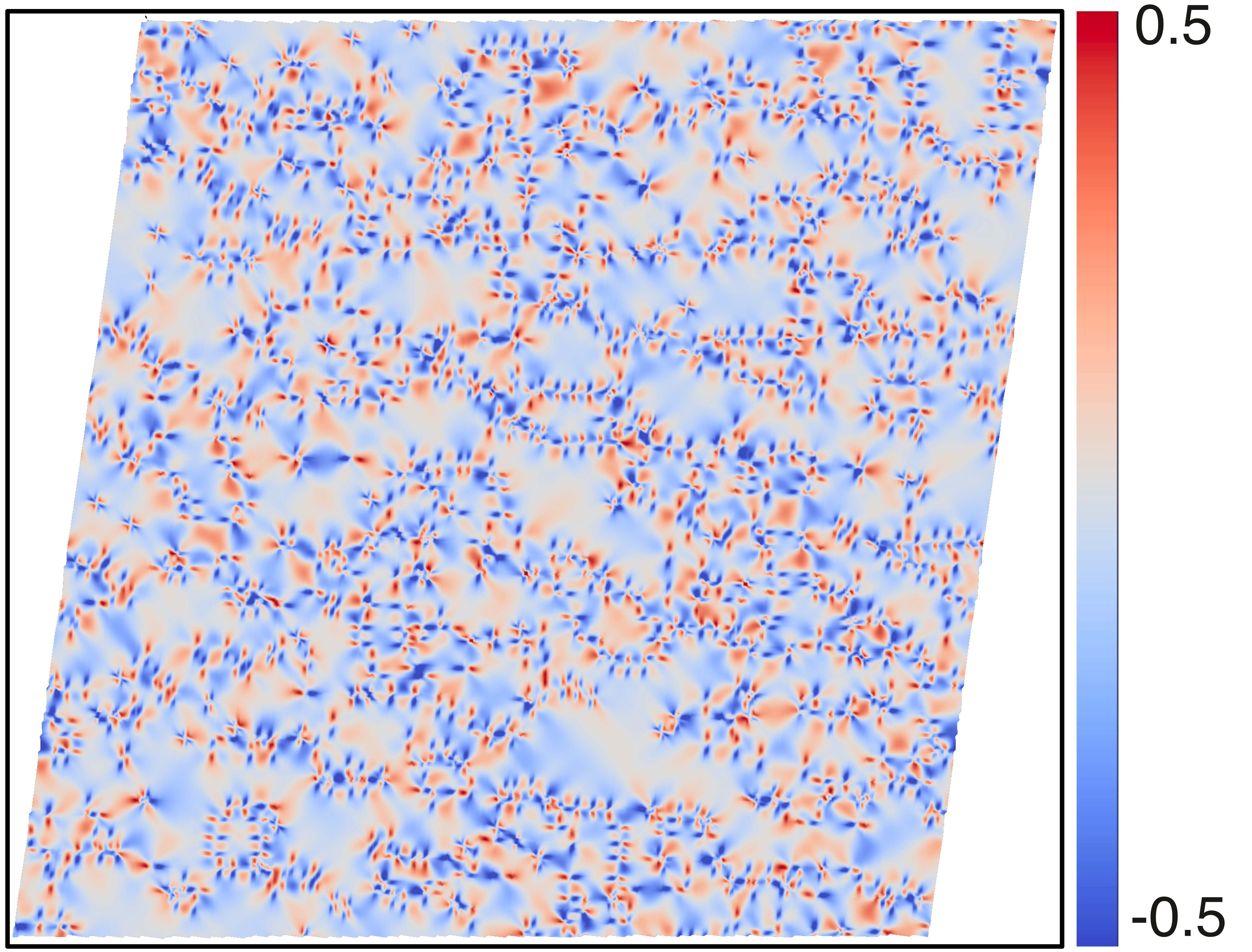}
\caption{Dislocated configuration emerging after a  homogeneous lattice is mechanically driven to the threshold of elastic instability. Colors indicate the level of the $\sigma_{xy}$ component of the Cauchy stress tensor. The highest stress levels of both signs correspond to the location of dislocation cores. Here $N=400$.} 
\label{fig:22}
\end{figure}

To obtain   a 'generic' sample, we first loaded a pristine (defect-free) crystal till the point of elastic instability of the affine configuration. The recorded  critical value of the loading parameter $\alpha=\alpha_c\approx0.138$ was found to be very close to the threshold of  positive definiteness of the acoustic tensor \cite{
%Simpson1989,
Grabovsky2013},
%,Steigmann2023
 see  the explicit formulas in \cite{SM}. To trigger the a generic instability we imposed a small Gaussian quenched disorder which was subsequently  removed. The breakdown of an elastic state took the form of  a massive  nucleation of dislocations  resulting in a  highly inhomogeneous configuration of the type shown in Fig.~\ref{fig:22}.   It is natural to assume that the  unfolding  of the elastic instability  was   arrested at a  state which is  only marginally stable and is therefore far from being random.

%\emph{Mechanical response. }
\begin{figure}[h!]

\includegraphics[scale=.113]{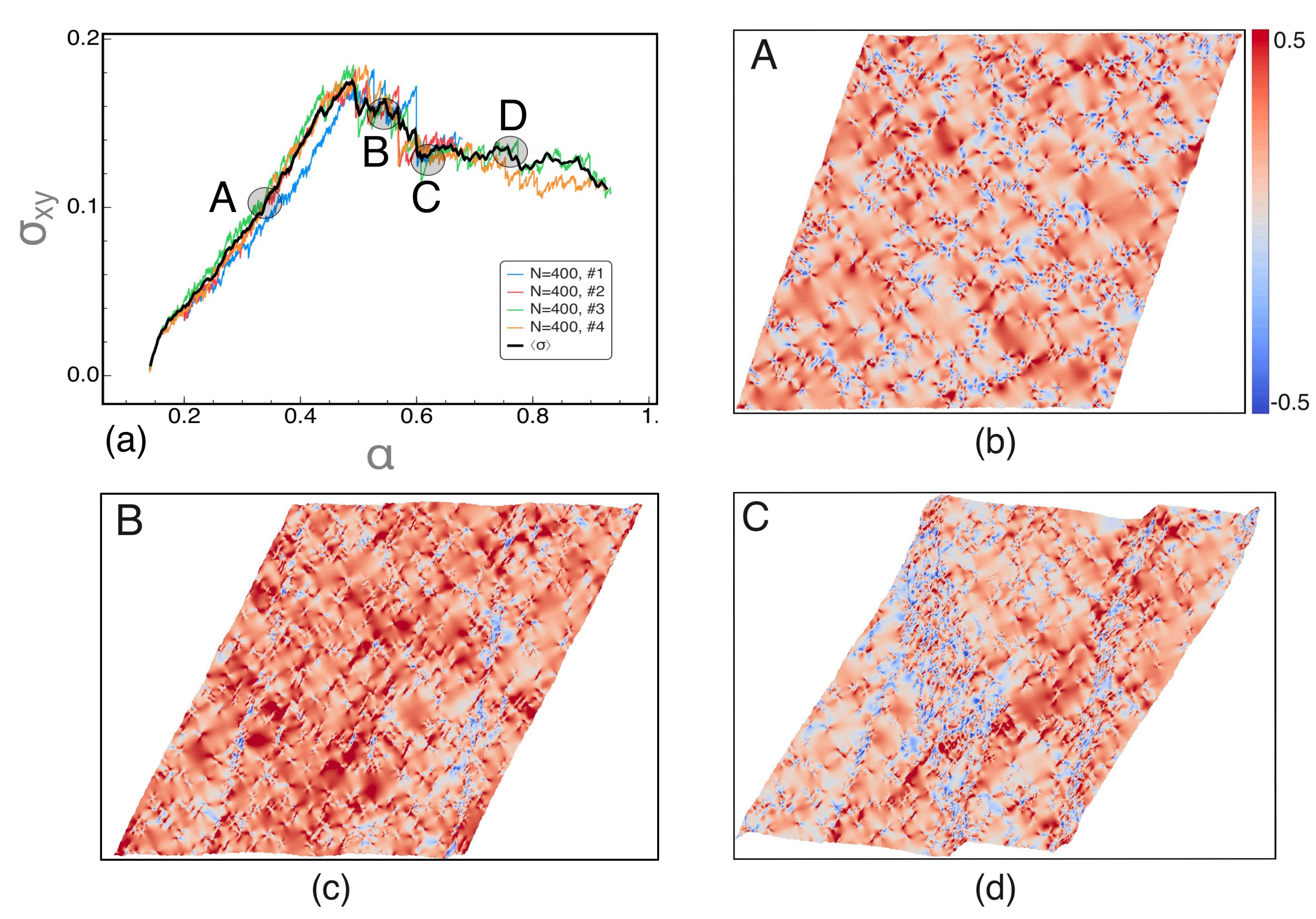}
\caption{(a) Stress-strain responses of four different   'generic' quasi-amorphous crystals of the type  shown in Fig. \ref{fig:22}; solid black curve indicates the average response. The stress-strain curves do  not start from the origin because    the global instability   does not relaxes stresses to zero  due to the emergence of  self-equilibrated (residual)  dislocation configuration. (b) Dislocated configuration   in a typical pre-yield state characterizing  the stage of 'microplasticity'.  (c,d) Dislocated configuration characterizing  different stages of   unfolding of the quasi-brittle system size event.
 Colors indicate the level of the $\sigma_{xy}$ component of the Cauchy stress tensor. In (a-d) N=400.
%The hardening stage from $D$ to $F$ shows the pre-(second) yield 'microplastic' deformation. The transition from $F$ to $H$ correspond to a cluster of system spanning avalanches. The post-(second) yield stationary plasticity regime is represented by the state $I$ on a stress plateau.
}
\label{fig:2}
\end{figure}

To study the mechanical  response of the obtained  'generic' sample we continued to load  it  quasistatically  while recording  the ensuing stress-strain relation.   The results are shown  in Fig.~\ref{fig:2}(a) for four  different realization of disorder with the solid black curve indicating the average response. Its salient feature is the presence of irregularly placed elastic branches interrupted by intermittent stress drops; the latter represent  plastic avalanches  involving  partial  unlocking of dislocation structures with superimposed  distributed nucleation/anihilation of dislocations.  A major diffuse stress drop, which we interpret as the  indication of the beginning of plastic yield,  separates two markedly different regimes. While the  quasi-elastic  pre-yield regime is characterized by  finite overall rigidity,  the  post-yield regime is represented by  an effective  stress plateau;  in both regimes we observe intense plastic fluctuations,  see Fig.~\ref{fig:2}(a).

%stands out and can be interpreted as a plastic  yield separating  

Our  Fig.~\ref{fig:2}(b) illustrates  the stress field in a typical pre-yield state $A$.  Comparing to Fig.~ \ref{fig:22}, the overall stress level is higher but the restructuring of the dislocation pattern  is relatively minor with dislocations mostly displaced   between the  preexisting locking sites. Such   'microplasticity'   response  is   a characteristic  pre-yield feature  for  both amorphous  and (defective) crystalline  solids \cite{Maas2018-qu,Papanikolaou2017-ld,
%Sparks2018-zp,
sparks2019avalanche,Rizzardi2022-iz}
%Duan2023-ue}. 
This   regime  ends with the  system-size  event which  takes the form of  single quasi-brittle plastic stress drop  in  well-annealed glasses and   some sub-micron crystals \cite{
%bei2008effects,chrobak2011deconfinement,wang2012pristine,
Cui2017-xn,Greer2011-hi,
%Brenner1956-rx,Brenner1958-dr,
Sharma2018-iw,Mordehai2018-qm}. 
%,Lilleodden2006-jq,Corcoran1997-vt}. 
 In   Fig.~\ref{fig:2}(c,d) we show the snapshots of the spatial  stress  configurations in the intermediate state $B$ and the  final state $C$ ending the transition to post-yield  regime. The comparison of Fig.~\ref{fig:2}(b-d) shows that the  associated  global restructuring involves collective dislocation activity leading to   the formation of  system-spanning shear bands.

\begin{figure}[h!]
\includegraphics[scale=.05]{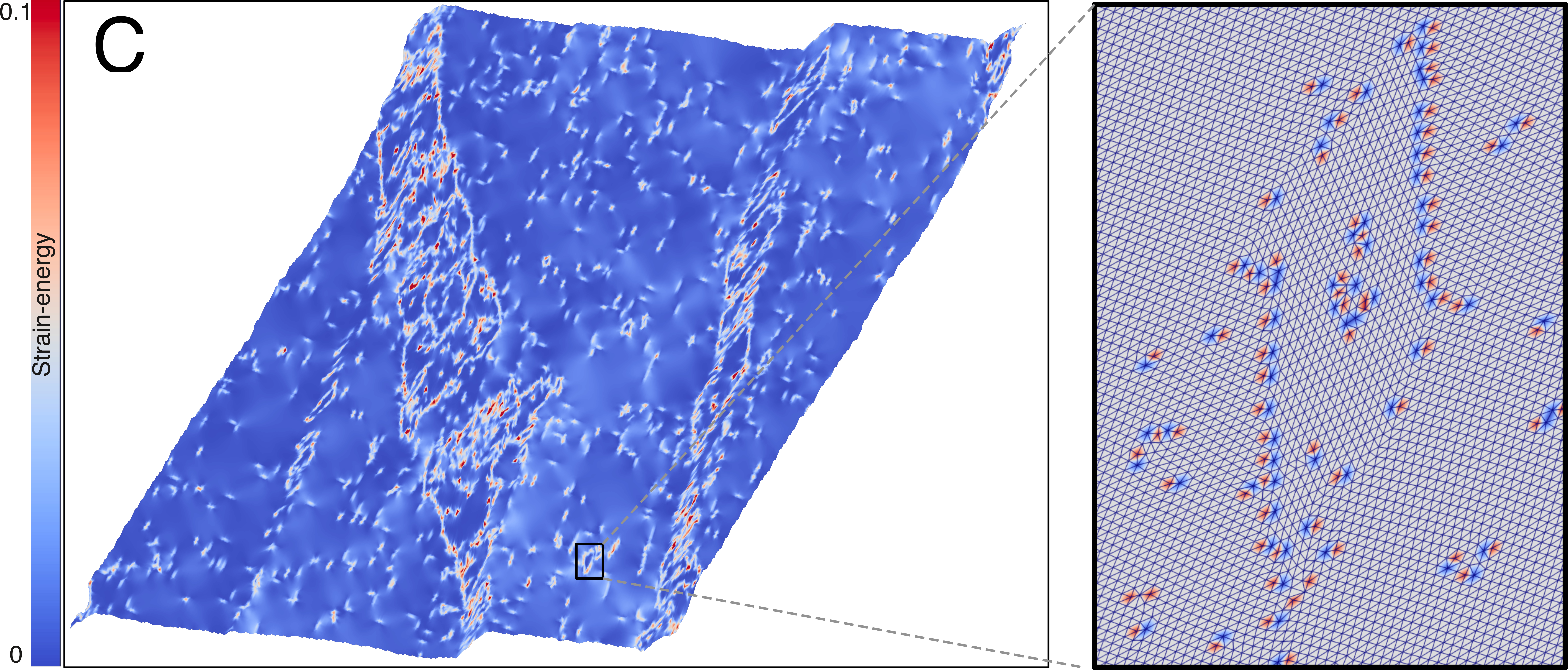}
\caption{The micro-configuration  configuration corresponding to the state $C$ marked  in Fig. \ref{fig:2}. Colors indicate the magnitude of strain-energy  density. The inset shows a single grain and details the dislocation structure of the grain boundaries. Here N=400.}
\label{grains3}
\end{figure}

 Additional  aspects of the state $C$ can be seen   in Fig.~\ref{grains3} where instead of the stress field, as in our Fig.~\ref{fig:2}(d), we now show  the strain-energy density field. In this representation the locations of individual dislocations become visible. The general pattern is the  development of  low-energy  patches of the original lattice forming polycrystalline grain texture with elastic energy   localized around  dislocation-rich grain boundaries.   In the inset presented  in Fig.~\ref{grains3}  we show  the  zoom in on the configuration of  elastic elements; individual dislocations   were identified via Delaunay triangulation  and  the  nodes with five/seven  neighbors  were shown in blue/red.  In particular, we see    the 
% emergence 
% of energy minimizing structure of grain boundaries 
ubiquitous presence  of   $\Sigma 5$ grain boundaries which can be explained by the fact that   
 the  misorientation of neighboring grains  is controlled by the overall compatibility of the deformation field \cite{Baggio2023-qu}. 
  
%  In the inset in Fig.~\ref{grains3}  we made  dislocations of different sign clearly visible: they were identified via Delaunay triangulation and   nodes with five/seven  neighbors  are shown in blue/red.  One can see    the emergence of energy minimizing structure of grain boundaries \cite{Sutton1984-us,Priester2012-wx,Randle2024-tf} with  ubiquitous presence  of   $\Sigma 5$ grain boundaries, see  \cite{Baggio2023-qu} for the explanation. 
% and dislocations correspond to adjacent pairs where the ordering 5-7 or 7-5 determines the sign).

 % It also shows  that the shift from 'local' to 'global' dislocation rearrangements leads to

% In other words, dislocations appear to be confined inside the effective cages where they can pile up but from where they can only rarely escape by breaking the existing locks.

%\begin{figure}[h!]
%%\includegraphics[scale=.1]{figures_ordering/figure_11.pdf}
%\includegraphics[scale=.053]{./Figures_ordered/Fig5.pdf}
%\caption{ Dislocated configuration   in a typical pre-yield state characterizing  the stage of 'microplasticity'. The state $A$  is marked  in Fig. \ref{fig:2}. Colors indicate the level of the shear component of the Cauchy stress tensor $\sigma_{xy}$.}
%\label{fig:031}
%\end{figure}

%It is natural to interpret the  regime-separating  system-size event  as a quasi-brittle plastic yield.

%  \begin{figure}[h!]
%\centering
%\subfigure[]{\includegraphics[scale=.0371]{./Figures_ordered/Fig6-a.pdf}\label{fig:03a}}
%\subfigure[]{\includegraphics[scale=.0371]{./Figures_ordered/Fig6-b.pdf}\label{fig:03b}}
%\caption{Dislocated configuration characterizing  different stages of   unfolding of the quasi-brittle system size event. The states $B$ and $C$ are marked  in Fig. \ref{fig:2}. 
%Colors indicate the level of the shear component of the Cauchy stress tensor $\sigma_{xy}$.
%} \label{fig:03}
%\end{figure}

\begin{figure}[h!]
\centering
\begin{tabular}{cc}
\includegraphics[height=3.2cm]{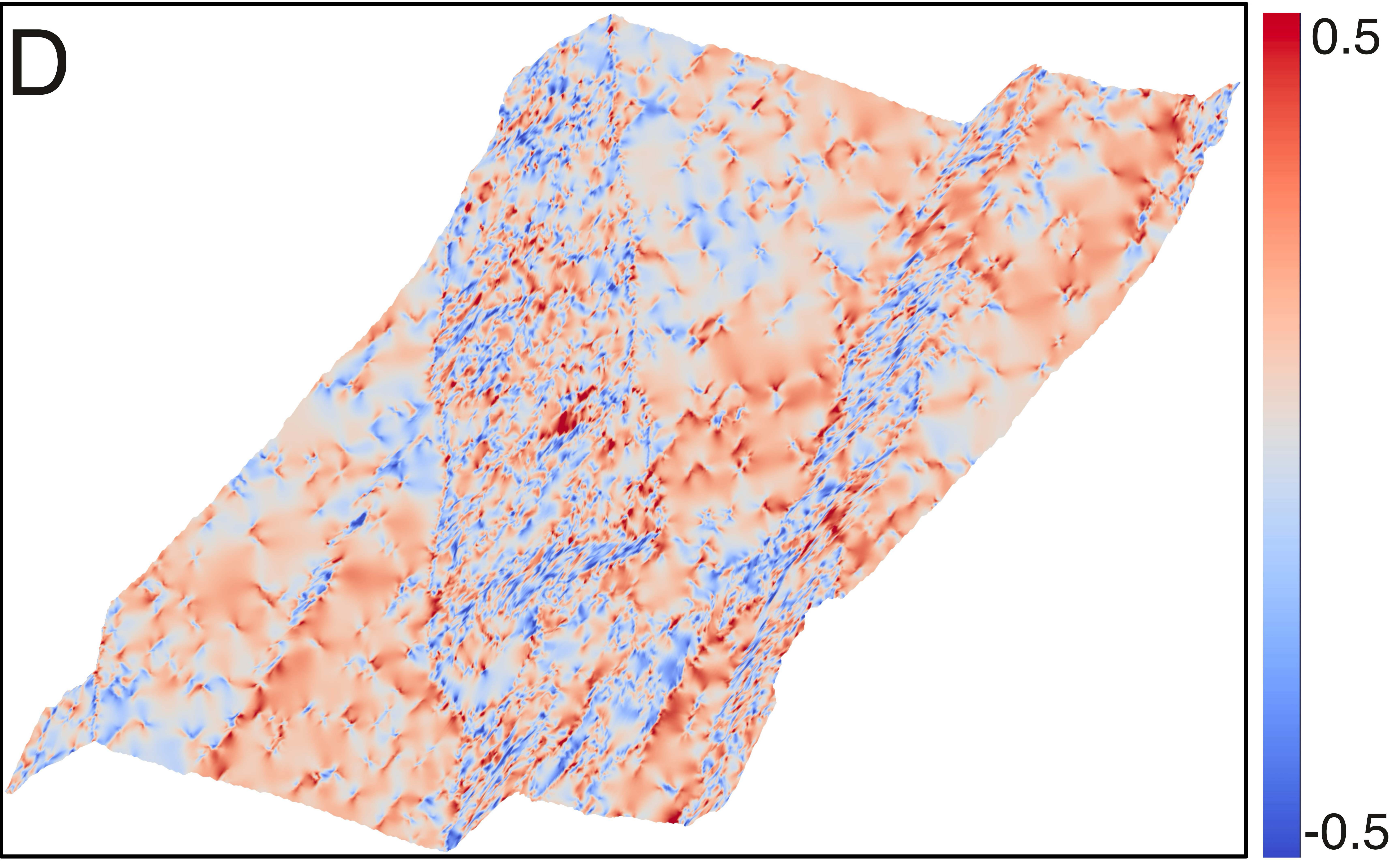} &
\includegraphics[height=3.2cm]{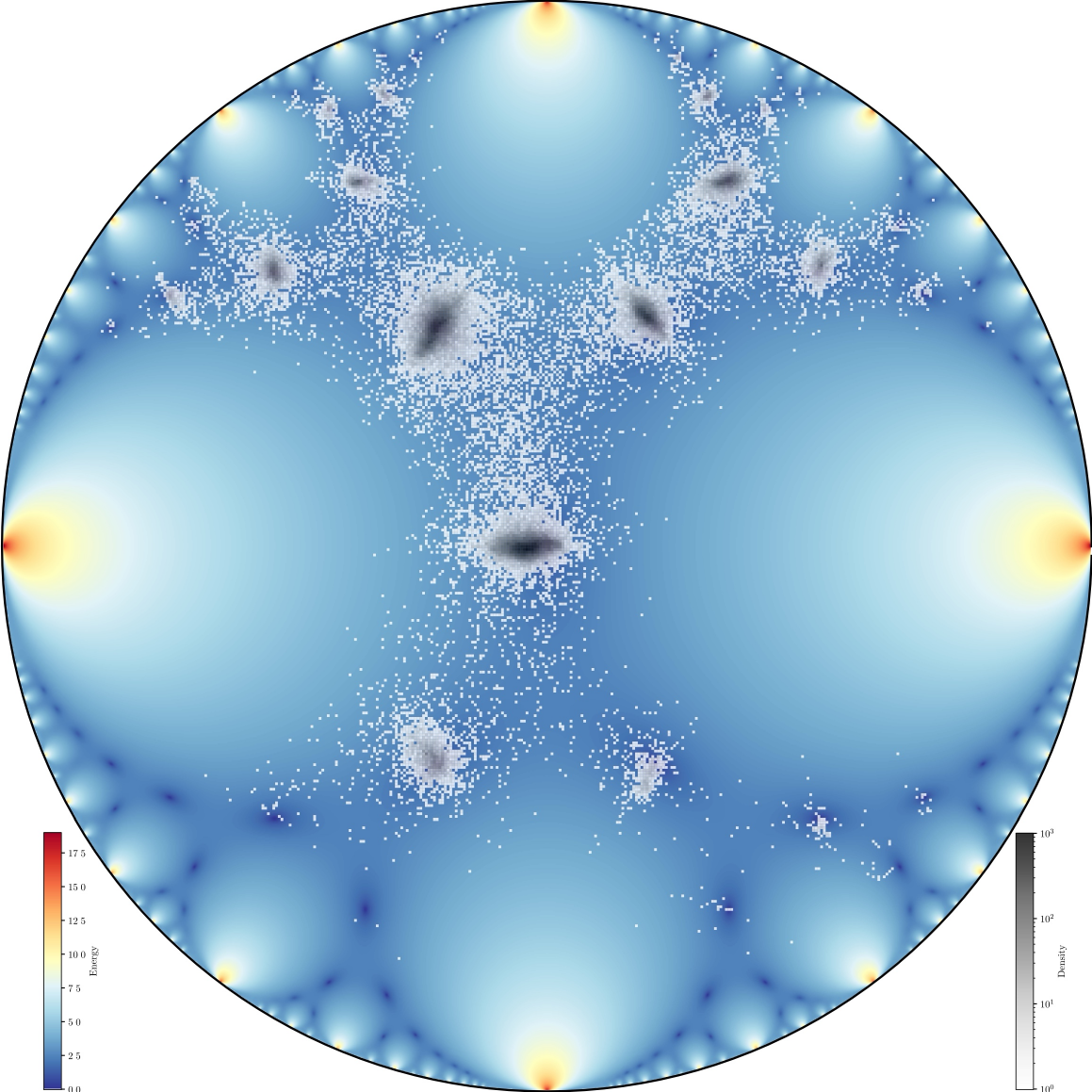} \\
(a) & (b)
\end{tabular}
\caption{(a) Shear band development in the regime of stationary post-yield plastic flow represented by the state $D$ in Fig. \ref{fig:2}. Here N=400. Colors indicate the level of the shear component of the Cauchy stress tensor $\sigma_{xy}$. (b) Corresponding distribution of local values of the metric tensors in the configurational space projected on the Poincaré disk. Background color  represents  the energy landscape whose 3D representation was  shown in Fig. \ref{fig:atlas0}(b). The color of the gray dots represents the number of the finite elements  sharing  the  same metric tensor.}
\label{fig:04}
\end{figure}

The ensuing post-yield regime,  characterized  by the relative stabilization of the average stress level, can be broadly interpreted as a quasi-stationary non-equilibrium steady state. A representative snapshot of the micro-configuration of the crystal in the state $D$ is shown in Fig.~\ref{fig:04}(a) where we see progressive maturation of the polycrystalline grain texture. During  intermittent avalanches characterizing this regime  larger  grains occasionally merge  while smaller grains continue to emerge. In this way the   self-organization of dislocations  continues to generate  additional scales which contribute to the  formation of the global hierarchical structure.  The complexity of the emerging   configuration  is  illustrated  in Fig.~\ref{fig:04}(b) where we show  the corresponding strain distribution in the  Poincaré disk. While in the reference state (pristine crystal) all configurational points were concentrated in the origin, now they are spread all over the configurational space with  expected accumulation near the bottoms of  at least five energy wells representing different replicas of the unstressed square lattice;  for comparison, in  the pre-yield  regime  the system  explores  at most  three energy wells immediately adjacent to the reference state. In other words,  plastic yield not only marks the transitions from isolated dislocation motion to collective dislocation behavior, but also  implies the  access to a much broader repertoire of relaxation mechanisms.  For instance, a more detailed analysis of the avalanche structure illustrated in \cite{SM} shows that while in the pre-yield regime plastification takes place  primarily in the form of isolated, linear arrangements of transformed elements,  a typical post-yield avalanche reveals extended plastified regions with complex branching.

An important window into  the  micro-mechanics  of solids  undergoing plastic flow is provided by the study of the statistical structure  of plastic fluctuations \cite{Papanikolaou2017-ld,Weiss2015-eh,
%Zaiser2006-gk,
%Tarp2014-ro,
%Miguel2006-wd,
Alava2014}.
%,Sethna2017}. 
In our  case  the relevant observables would be  the energy dissipated at a single avalanche  $\Delta W$ and the corresponding macroscopic stress drop $ \Delta\Sigma= A \Delta\sigma_{xy}$, where $A$ is the surface area of the sample. It is known that in amorphous solids the probability distribution  of these observables is compatible with   the finite-size scaling  ansatz 
%\begin{equation}
 $ P(s; L) \;=\; s^{-x}\, \mathcal{F}\!\left(\frac{s}{L^{D_x}}\right),$
%\end{equation}
where $L=Nh$  is the linear system size, $h$ is the size of an element, $x$ is the power-law exponent  with $x = \tau$ for stress drops and $x = \epsilon$ for dissipated energy and $D_x$ is the corresponding fractal dimension ($D_\epsilon$ or $D_\tau$, respectively). Finally  $\mathcal{F}(u)$ is a universal scaling function. 
It is also known that in amorphous plasticity  the  exponents $\epsilon$ and $\tau$ are similar  for  pre-yield and post-yield regimes while the exponents $D_\epsilon$ and  $D_\tau$ may be different   \cite{Tyukodi2016-dv,Budrikis2017-ex,Lerner2018-ue,Ozawa2022-xb}. 

\begin{figure}[hbt!]
\centering
\vspace{3mm}
\begin{tabular}{cc}
\includegraphics[width=0.235\textwidth]{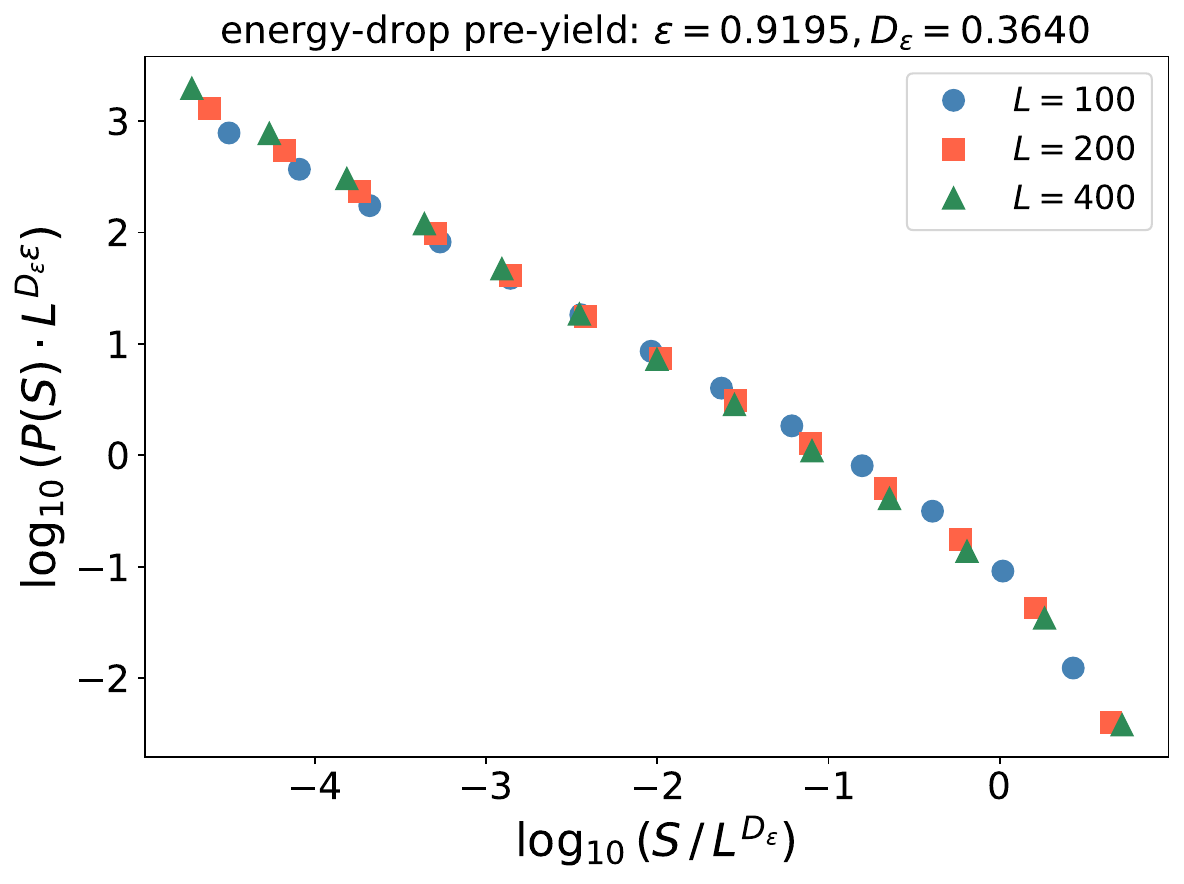} &
\includegraphics[width=0.235\textwidth]{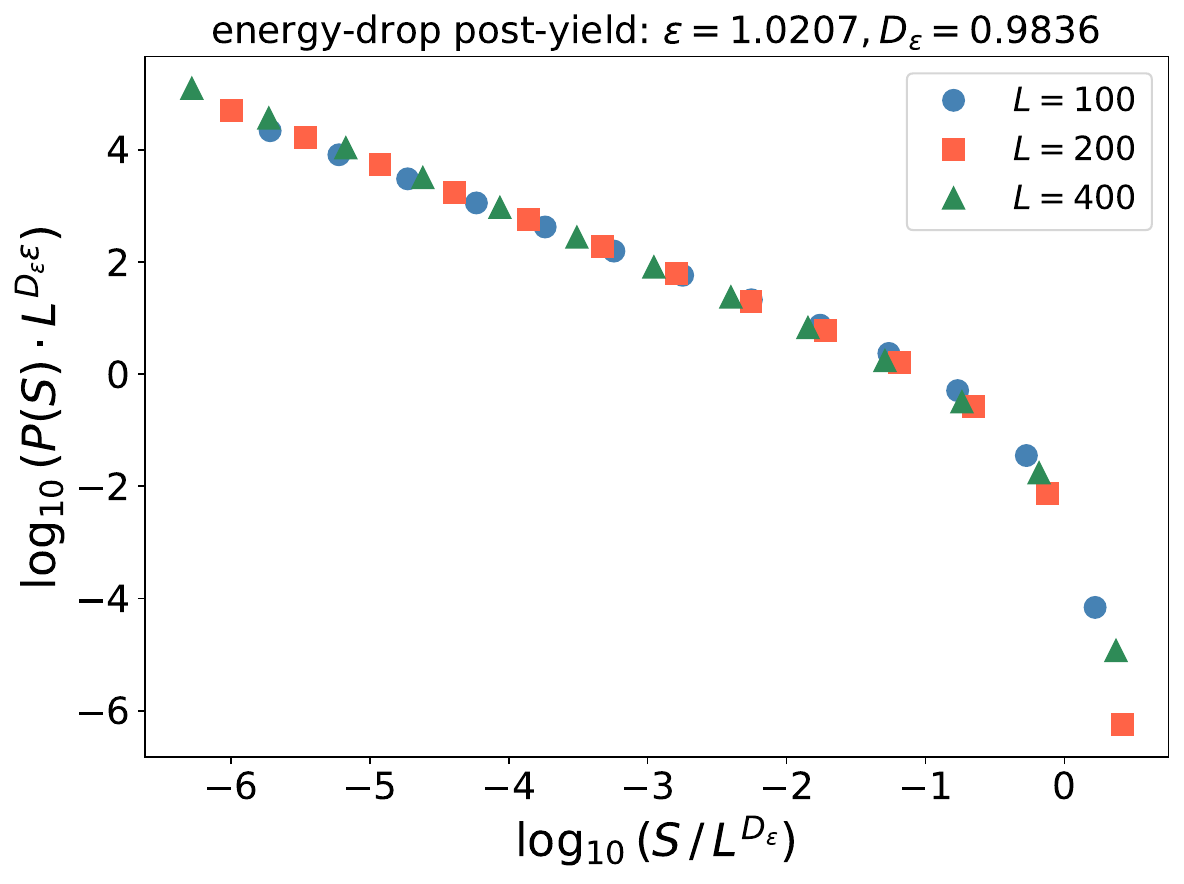} \\
(a) 
%Pre-yield (Energy) 
& (b) 
%Post-yield (Energy) 
\\
\includegraphics[width=0.235\textwidth]{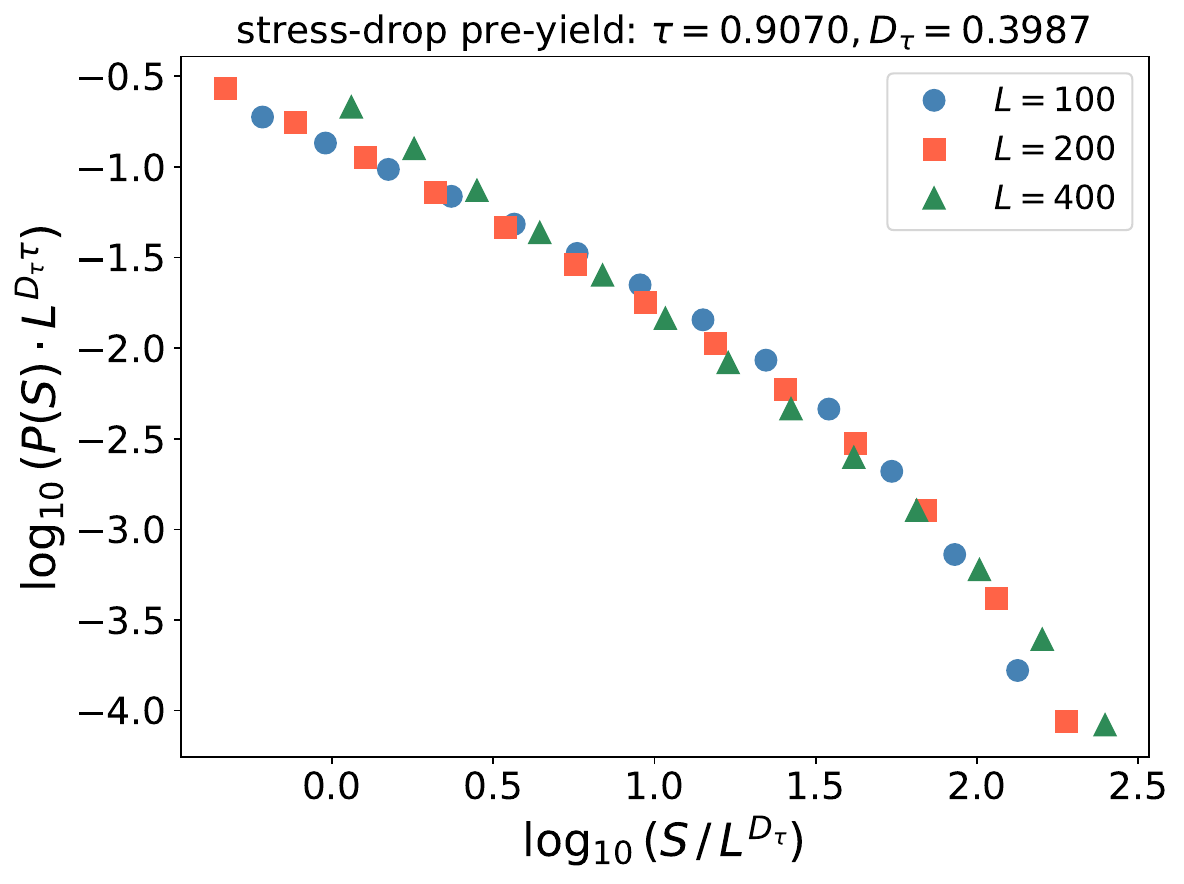} &
\includegraphics[width=0.235\textwidth]{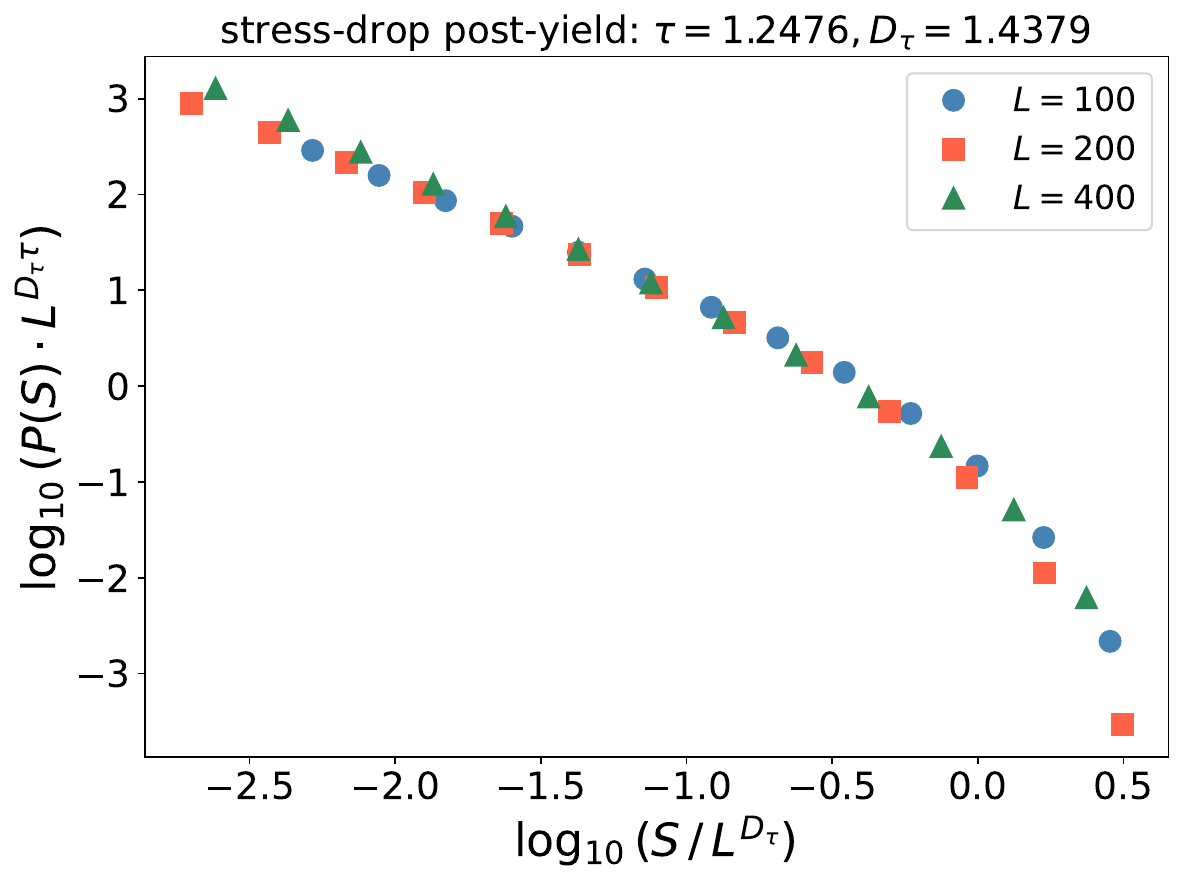} \\
(c) 
%Pre-yield (Stress) 
& 
(d) 
%Post-yield (Stress) 
\\
\end{tabular}
\caption{Optimal data collapses of the probability distributions onto universal master curves. (a, b) Collapse of the energy drops $\Delta W$ in the pre-yield  generating  exponents $\epsilon=0.9195$ and $D_\epsilon = 0.3640$  (a)  and post-yield regimes generating  exponents $\epsilon=1.0207$ and $D_\epsilon=0.9836$ (b).  Collapse of the  macroscopic stress drops $\Delta \Sigma $   in the pre-yield generating  exponents $\tau=0.9070$ and $D_\tau=0.3987$ (c) and   post-yield regimes,  generating  exponents $\tau=1.2476$ and $D_\tau=1.4379$ (d).   
  }  
\label{sm:fig_collapses}
\end{figure}

To determine  these power-law exponents for  our numerical experiments   we employed  a robust data-collapse optimization procedure detailed in \cite{SM}.  The resulting data collapse is illustrated in Fig.~ \ref{sm:fig_collapses}. One can see  that   for  our 'generic' quasi-amorphous crystals  
the statistics of   avalanches follows the same pattern as for their amorphous counterparts. In particular, similar to the case of well-annealed glasses  \cite{Papanikolaou2017-ld,Weiss2021-tt,Sethna2001,Ispanovity2014-ra,Alava2014,Lehtinen2016-qy},  the  scaling behavior  remains basically unchanged across the  quasi-brittle yielding transition. 

More specifically,  the  obtained   power law range  for energy drops extends over up to six decades with  pre-yield and post-yield energy distributions sharing almost the same  value of the exponent  $\epsilon = 1.0$. The invariance of this exponent   across the yielding transition  suggests that while the spatial organization of plastic events changes drastically   from local and non-cooperative to global and collective, the statistical distribution of   energy drops    remains insensitive to these  details responding only to the global  organization of the underlying energy landscape \cite{Shang2020,oyama2021-rf}.

%\begin{figure}[h!]
%\includegraphics[scale=.1]{./Figures/Fig7.pdf}
%\caption{Probability distribution of energy drops   during avalanches in pre- and post-yield  regimes. 
%%Both distributions are well aprroximated by a power-law with exponential cutoff.  The power law exponents $\epsilon$ in both regimes are identical. Instead the  cutoff parameters  $\lambda$  are drastically different. 
%}\label{fig:energy_drops}
%\end{figure}

The value of the power-law exponent $\epsilon \approx 1.0$ is a signature of an archetypically 'wild' crystal plasticity in the sense of \cite{Weiss2015-eh}. The same value was obtained in discrete dislocation dynamics studies of disorder-free 2D samples containing a fixed number of preexisting  dislocations \cite{Ispanovity2014-ra,PhysRevMaterials.5.073601} and was also recorded in numerical experiments on pristine crystals using a scalar version of the MTM \cite{Zhang2020-ax}. The exponent $\epsilon \approx 1.0$ has been previously interpreted in crystal plasticity framework as representing either dislocation jamming or self-induced glassiness \cite{Ovaska2015-yb,Lehtinen2016-qy,Zhang2016-gh}. 

It is rather striking that exactly the same value of the exponent has been also recorded in the studies of plastic flows in structural glasses~\cite{tyukodi2019avalanches,Ferrero2019-rx,Shang2020}. A theoretical understanding of the relation between the emergence of the power law exponent   $\epsilon\approx1.0$ and the structure of the underlying energy landscapes has been developed in the theory of spin glasses. Thus,  it was shown that the hierarchical (ultrametric) organization of energy wells in the phase space results in an intermittent, scale-free response to quasistatic deformation ~\cite{Franz2017-fa,Muller2015,Berthier2019-dm,Dennis2020-jb}. Moreover, it was rigorously proved   that in the mean-field limit the associated distribution of energy avalanches must be of a power law with exponent $\epsilon \approx 1.0$. In the spin glass context it has been also understood that the reason behind this particular value of the exponent is the marginal stability of the system ~\cite{Franz2017-fa,Pazmandi1999-lb,Le-Doussal2012-kv}.

Since glasses can be viewed as liquids at a quenching threshold, their marginal stability  is  similar to marginality  exhibited by  granular matter at a jamming point. If such  systems are mechanically driven, the proximity to unstable modes  leads to the mixing of statics (stability) and dynamics (instability) making the  mechanical response inherently  intermittent ~\cite{lamp2022brittle, rossi2022emergence}. Similarly, since  our 'quasi-amorphous crystal' emerged from arrested dynamics, one can argue that the associated  self-generated disorder   brought the system  from a solid to  an effectively  glassy state.  Furthermore, our results  suggest that the implied  marginality is  not   affected by the quasi-brittle yield,  which apparently does not compromise the global organization of the energy landscape while, of course,  affecting the reachability of its different subdomains~\cite{
%Regev2017-ks,Regev2017-jr,
Regev2021-qf}.

The situation is not exactly the same for the exponent $\tau$ characterizing the power law statistics of the stress drops $\Delta \Sigma$. Note first that in the pre-yield regime, the power-law exponents for stress and energy are rather  similar, $\tau \approx \epsilon$,  which  suggests that the dissipated energies and the stress drops are roughly   proportional, $\Delta W \sim \Delta \Sigma$. Such linear scaling  implies some kind of caging,  which means  a highly constrained regime of dislocation motion with  individual avalanches arrested  by a regular arrays of obstacles. This may be also the reason behind the strong  hardening behavior in the pre-yield regime.

Instead, in the post-yield regime the recovered  exponents $\tau$ and $\epsilon$ are markedly different. Here 
individual avalanches  are no longer hindered by local structural constraints and one can assume that the stress drop is only constrained by the corresponding  strain increment due to the constraints of (nonlinear) elasticity. Then, if we assume that $\Delta W \sim \Delta \Sigma ^\gamma$ we can expect that $\epsilon - 1 = (\tau - 1)/\gamma$.  In particular, if  elasticity is linear, $\gamma=2$, and the measured   value is $\tau \approx 1.2$, we can expect that $\epsilon \approx 1.10$ which  is in reasonable agreement with our  measured value  $\epsilon \approx 1.02$ given that the uncertainty  for  our relatively small system  is of the order $\sim \pm 0.2$, see \cite{SM} for details.

Note next that   the pre-yield  values of the cutoff parameters $L^{D_x}$ are  much smaller than  their  post-yield  values which    indicates that larger avalanches  become  more probable ~\cite{Lin2014-bx,ruscher2021avalanches}. The emergent collective behavior,  reflected in  such  broader extent of avalanche activity, reaches  in   post yield regime  the size of the system  suggesting the divergence of the characteristic length.   Further differences between pre-yield and post-yield  mechanical responses emerge from the analysis  of the  values of the exponents  $D_x$  providing   information  about the    morphology of the plastically deforming regions. Thus, in In the   pre-yield regime   the energy avalanches  are characterized by  a low fractal dimension  $D_\epsilon \approx 0.36$, pointing towards highly localized, spatially scattered, macroscopically  isolated rearrangements. Instead, in the  post-yield regime we obtain the value  $D_\epsilon \approx 0.98$ which suggests the emergence of 1D-like  system spanning bands. The latter  reflect  highly  cooperative nature of the underlying plastic deformation and the physical nature of the implied  criticality  is presently actively  debated \cite{Grinstein1995-lw,Zippelius2023-ak,
%Charbonneau2023-oo,
Oyama2023-oj,Xing2024-tz,Zhang2020-ax}.

%
% , extended nature of the individual yielding events dominating the mature flowing stage, where typical avalanches include the motion of grain boundaries whose lengths are comparable to linear system, see Fig.~\ref{sm:fig_plastified}(b).
%
%
%
%
% Focusing specifically on the dissipated energy avalanches, we find a stark contrast in the measured fractal dimensions between the two deformation stages. 
%
%
%
%
%
%, see Fig.~\ref{sm:fig_plastified}(a).
%
%
%
%

%
% which implies that  the system was stabilized close to the associated threshold. 
%
%We can then argue that the outcome of  our 'proparation' protocol was 
%
% Since we observe the same 
%The fact that the value of the exponent $\epsilon = 1.0$ was observed in both pre- and post-yield regimes, 

%This conjecture is supported  by the observation  that  shear bands  in the post-yield regime   traverse the whole computational domain which is in   contrast to the  subextensive localized dislocation rearrangements  in the pre-yield regime.

%Finally we recall  that  intermittent fluctuations in amorphous plasticity are usually modeled in terms of 

Finally we mention that the observed similarity between the  structure of  intermittent plastic fluctuations in amorphous and crystalline solids is supported by the fact that elasto-plastic models, typically used to model amorphous plasticity  \cite{Nicolas2018-iy,PhysRevLett.129.228002,Ferrero2019-rx,Fernandez-Castellanos2021-yn}, are very similar to the MTM  which operates within basically the same finite element setting. The difference is that the phenomenological yield thresholds of elasto-plastic models are replaced in the MTM by elastic instabilities originating from the non-convexity of the  energy landscape. Accordingly,  the fixed linear elastic propagators are replaced by the solution 'on the fly' of the corresponding nonlinear elasticity problems. Note also   that the crucial   nonlinearity in the MTM is  of universal geometrical nature as it originates from the exact description of finite elastic deformations. However, the observed general agreement in the values of the computed exponents suggests that we are dealing with the same universality class.

% 
%To conclude,  we used a novel mesoscopic tensorial model to achieve a deeper understanding of the phenomenon of plastic yield in crystals in the limit of zero strain rate and negligible thermal effects.

 To conclude,  we  showed that mechanically driven perfect  crystals can exhibit   quasi-brittle plastic yielding which is remarkably similar  to the behavior  of well-annealed glassy materials. The implied parallel features of the mechanical response emerge after pristine crystals  acquire self-generated disorder  resulting  from  massive dislocation nucleation  during  the breakdown of an affine elastic configuration.  The ensuing quasi-amorphous crystals  exhibit  pre- and post-yield avalanches with  power law statistics whose  matching exponents are indicative of  the marginality  of the associated mechanical system.  Adaptation of the same  model to realistic 3D crystals will allow one to distinguish  between edge and screw dislocations opening the way to the adequate account  of such physically important  effects as dislocation climb, cross slip and forest hardening. To be adequate such an analysis should be, of course,  performed with the account of  the effects of finite  temperature   \cite{Zhang2023-am, Li2024-ax}.
 
% introduces a timescale causing "thermal rounding" of stress-strain peaks [Zhang et al., Acta Mater. 260, 119201 (2023); Li et al., arXiv:2409.16987 (2024)]. Our athermal limit represents the limit-envelope of these processes, where the universal scaling $\epsilon \approx 1.0$ is expected to persist despite mechanisms like climb becoming active at higher temperatures.}
%
% 
% It would  also be  also of interest to explore whether crystals with higher triangular symmetry exhibit similar behavior and to assess whether the specific choice of strain-energy density functional plays a crucial role in these phenomena. 

 %with HCP, FCC and BCC symmetry. In particular, only a 3D model will 
% 
 \emph{ Acknowledgments.} The authors thank N. Gorbushin   for helpful comments. The work of O.U.S. was supported  by  the grants   ANR–21-CE08-MESOCRYSP, ANR-20-CE91-0010,  ANR-24-CE91-0002.   L. T.  was supported by the grants  ANR–21-CE08-MESOCRYSP and ERC-H2020-MSCA-RISE-2020-101008140.

%\bibliography{formatted}

\end{document}

\title{Self-induced marginality in plastically deformed crystals: Supplemental Material}
\author{O.U. Salman}
\affiliation{
LSPM, CNRS UPR3407, Université Sorbonne Paris Nord, 93400, Villateneuse, France}
\affiliation{
Lund University, Department of Mechanical Engineering Sciences, SE-221 00 Lund, Sweden}
\author{A. Ahadi}
\affiliation{
Lund University, Department of Mechanical Engineering Sciences, SE-221 00 Lund, Sweden}
\author{ L. Truskinovsky}
\affiliation{
PMMH, CNRS UMR 7636 ESPCI PSL, 10 Rue Vauquelin, 75005, Paris, France}

\date{\today}
 \maketitle

\paragraph {Fundamental and elastic domains.}  Consider  the energy density  function $ \phi({\bf C})$   with tensorial symmetry  $ GL(2,\mathbb{Z}) = \left\{{\bf m},\, m_{IJ}\in{\mathbb Z},\, \det({\bf m})=\pm 1 \right\}.$    We  refer to the restriction of such periodic  energy density 
to the minimal periodicity (fundamental) domain   as $\phi_\mathscr{D}(\tilde{\bf C})$. Here $\tilde{\bf C}={\bf m}^T{\bf C} {\bf m}$  is the projection of  a  general metric tensor ${\bf C}$ into the domain $\mathscr{D}$ and   ${\bf m}$ is the corresponding unimodular integer valued matrix  that performs the projection  
while  ensuring that $ \phi({\bf C})= \phi_\mathscr{D}(\tilde{\bf C})$.  The  actual  configuration  of the fundamental domain  $\mathscr{D}$ is well known   \cite{Parry1998,Conti2004-sv,Engel2012,pitteri2002continuum}
\begin{equation}
\mathscr{D} = \left\{  0<C_{11}\le C_{22},\quad 0\le C_{12}\le \frac{C_{11}}{2}\right\},
\end{equation}
where $ \det{\bf C}=1$. Given a generic metric $\bf C$, the task of finding a unimodular matrix ${\bf m} $ ensuring that $\tilde{\bf C}\in \mathscr{D}$ (and therefore $ \phi({\bf C})= \phi_\mathscr{D}(\tilde{\bf C})$) can be formulated as a recursive algorithm which is also well known  (Lagrange reduction) \cite{Conti2004-sv,Engel2012}. Specifically, if we define the matrices   
\begin{equation}
{\bf m}_1=\begin{pmatrix}
1 & 0 \\
0 & -1 
\end{pmatrix},\,\,\,
%\end{equation}
%\begin{equation}
{\bf m}_2=\begin{pmatrix}
0 & 1 \\
1 & 0 
\end{pmatrix},\,\,\,
%\end{equation}
%and  \begin{equation}
{\bf m}_3=\begin{pmatrix}
1 & -1 \\
0 & 1 
\end{pmatrix},
\end{equation} 
%we can formulate the reduction procedure in a form of an explicit algorithm. Thus, if we
 and start with an assumption that    ${\bf m} = \mathbb{I}$, we can  proceed  through the following steps: 
(i ) if $C_{12}<0$, change sign of $C_{12}$ using the mapping ${\bf m}\rightarrow ${\bf m}${\bf m}_1$; 
(ii) if $C_{22}<C_{11}$, swap these two components using the mapping ${\bf m}\rightarrow ${\bf m}${\bf m}_2$;
(iii) if $2C_{12}>C_{11}$, set $C_{12}=C_{12}-C_{11}$ using the mapping  ${\bf m}\rightarrow ${\bf m}${\bf m}_3$. 
Note  that  the action of   ${\bf m}_1$ is a reflection that ensures an acute angle between two lattice vectors of the square lattice $\mathbf{e}_i$ where $i=1,2$; the action of the matrix ${\bf m}_2$ is also a reflection as it swaps two lattice vectors $\mathbf{e}_i$. Both of these operations  belong to the point group and propagate the metric only inside the corresponding 'elastic  domain'  (Ericksen-Pitteri neighborhood)  composed of the four copies of the minimal  domain $\mathscr{D}$ ~\cite{pitteri2002continuum,Baggio2023}. Instead, the mapping defined by matrix ${\bf m}_3$ brings the metric outside the 'elastic domain'  and therefore represents a quantized analog of the macroscopic plastic strain.
\begin{figure}[h!]
\centering
\includegraphics[scale=0.25]{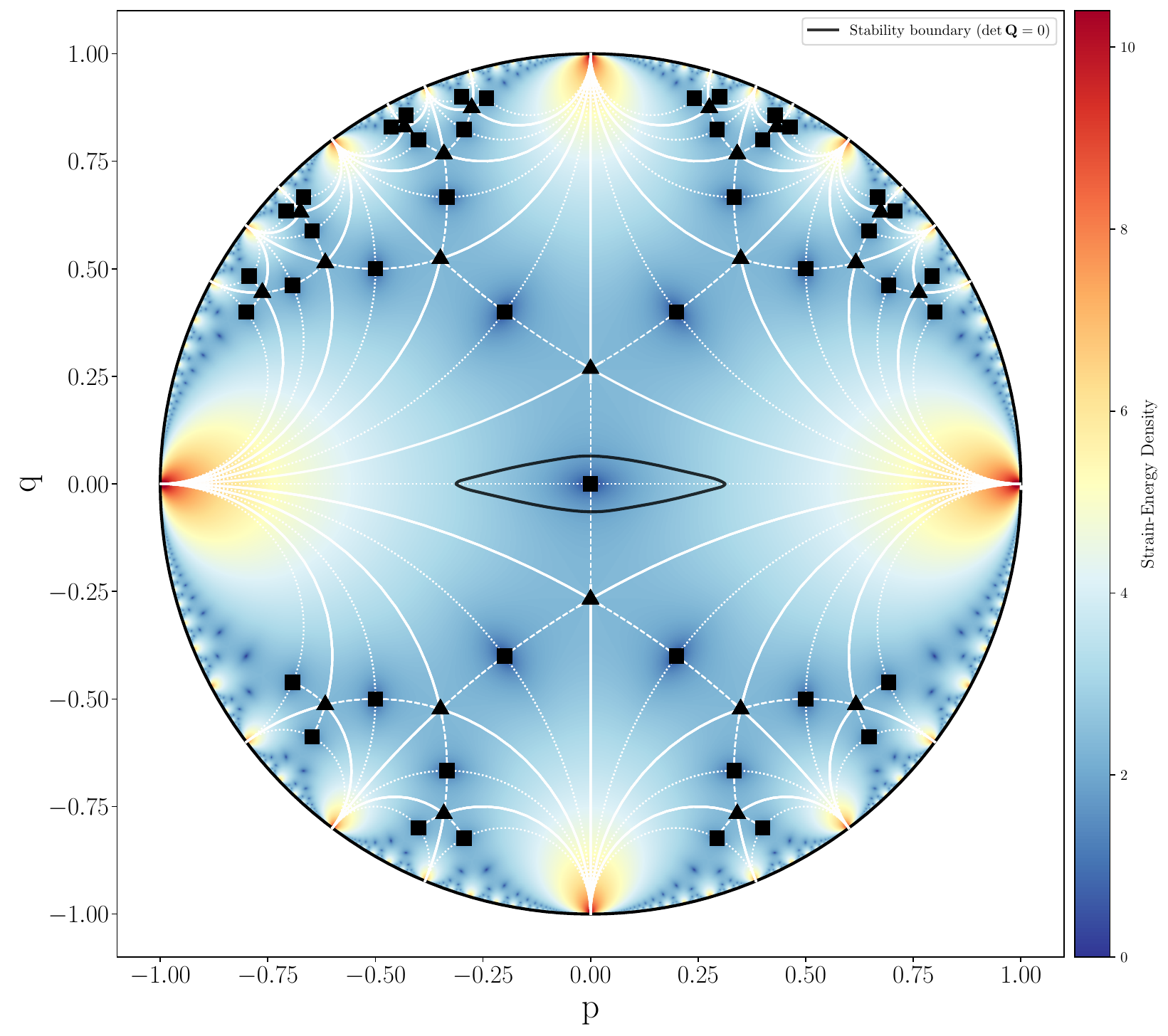}
\caption{Strain-energy density  in the configurational space $C_{11}, C_{22},C_{12}$ with  $\det{\bf C}=1$  stereographically projected on  the  Poincar\'e disk.
%  $p^2 + q^2 < 1$ where 
%$p = t(C_{11} - C_{22})/2$, 
%$q = tC_{12}$, 
%and  $t = 2/(2 + C_{11} + C_{22})$.
 Thin white lines represent  the $GL(2,\mathbb{Z})$ tessellation of the  a Poincar\'e disk  into equivalent periodicity domains. Colors represent energy levels following  Fig.~ \ref{fig:atlas0}(b) in the main text. Elastic stability boundary given by ~\eqref{detQ} is  shown by the thick black contour surrounding  the reference state.  Small black squares and triangles correspond to equivalent square and triangular lattices, respectively.
}
\label{01}
\end{figure}

\paragraph{Elastic energy density.} While the single-period Landau potential $\phi_\mathscr{D}(\tilde{\bf C})$ can be   constructed using the classical Cauchy-Born approach, see for instance   \cite{Baggio2023-qu, Baggio2023},  in this paper we used, for simplicity, the phenomenological expression proposed in \cite{Conti2004-sv}. Specifically, the  elastic energy density is represented as a sum of two terms:
 \begin{equation}
\phi(\tilde{\bf C}) = \phi_0 \left(\frac{\tilde{\bf C}}{(\det {\tilde{\bf C}})^{1/2}} \right) + \phi_v(\det \tilde{\bf C}), 
\end{equation}
where $\phi_0 $ accounts for contributions due to   shear   while $\phi_v$ penalizes    volumetric deformations. The $ GL(2,\mathbb{Z})$ invariant shear  term is chosen in the form  a sixth-order polynomial  which is the minimal requirement  ensuring   stress continuity over the whole configurational space  \cite{Parry1998,Conti2004-sv}. It depends on a single parameter $\beta$ allowing one to associate   ground state either with square or triangular (hexagonal) lattice 
 \begin{equation}\label{enerC}
 \phi_0 \left( \frac{\tilde{\bf C}}{(\det {\tilde{\bf C}})^{1/2}} \right) = \beta\psi_1\left( \frac{\tilde{\bf C}}{(\det {\tilde{\bf C}})^{1/2}} \right) + \psi_2\left( \frac{\tilde{\bf C}}{(\det {\tilde{\bf C}})^{1/2}} \right). 
 \end{equation}
The requirement of stress continuity across the whole periodic energy landscape specifies the functions $\psi_1$ and $ \psi_2$ 
\begin{align}
\psi_1 &= I_1^4 I_2 - \frac{41}{99} I_2^3 + \frac{7}{66} I_1 I_2 I_3 + \frac{1}{1056}I_3^2, \\
\psi_2 &= \frac{4}{11} I_2^3 + I_1^3 I_3 - \frac{8}{11} I_1 I_2 I_3 + \frac{17}{528} I_3^2.
\end{align}
where  we used  the known hexagonal invariants of the metric tensor 
\begin{align}
I_1 &= \frac{1}{3}(\tilde{C}_{11} + \tilde{C}_{22} - \tilde{C}_{12}), \\
I_2 &= \frac{1}{4}(\tilde{C}_{11} - \tilde{C}_{22})^2 \nonumber\\
    &\quad + \frac{1}{12}(\tilde{C}_{11} + \tilde{C}_{22} - 4\tilde{C}_{12})^2, \\
I_3 &= (\tilde{C}_{11} - \tilde{C}_{22})^2(\tilde{C}_{11} + \tilde{C}_{22} - 4\tilde{C}_{12}) \nonumber\\
    &\quad - \frac{1}{9}(\tilde{C}_{11} + \tilde{C}_{22} - 4\tilde{C}_{12})^3.
\end{align}
We select   $\beta = -1/4$ to   ensure  that global energy minimizers correspond to square lattices. 
 The volumetric part of the energy density, which primarily influences the fine structure of the dislocation cores   and also controls the formation of voids, is assumed to be of a generic form preventing infinite compression
 % \begin{equation}
$ \phi_v(s) = \mu(s - \log(s)). $
% \end{equation}
  To keep in our numerical experiments the strain field close to the surface $\det{\bf C} = 1$, we adopted   a sufficiently high value of the bulk modulus,  $\mu = 5$.

The isochoric part of the  energy landscape is illustrated in Fig. 2 of the main paper. Here we also show the same Landau-type potential in Fig. \ref{01}, but now in the level set representation. Dark blue regions  correspond to the equivalent square energy wells. Red domains  show effectively inaccessible regions where the energy is very high (truncated). Relatively low energy valleys colored in yellow correspond roughly to two simple shear plastic 'mechanisms' available in square lattices.

\paragraph{Internal length scale.} 
The lack of convexity of the potential is a property which  the MTM   shares with other similar Landau-type continuum theories. In the MTM  approach the necessary regularization  is  introduced by bringing  into the model   an internal  scale  $h$ through  explicit spatial discretization.    If $L$ is the linear size of the macroscopic domain and $N^2$ is the number  of  nodes in the mesoscopic  grid, then the parameter $h=L/N$  introduces elastic finite  elements imitating mesoscopic aggregates of atomic particles and defines in this way the  cutoff beyond which the deformation is considered homogeneous (affine).  The corresponding small parameter is $\delta=h/L=1/N$. Another small parameter in the problem is  $\epsilon=a/L$ where $a$ is  the   interatomic distance and we  are interested in the  double limit:  $\epsilon \to 0$  and $\delta \to 0$ while  $\delta\gg\epsilon$. In the MTM approach  we first implicitly perform the limit $\epsilon \to 0$ and recover in this way  the  continuum  constitutive response    using  the  Cauchy-Born rule. Then instead of performing the  classical  $\delta \to 0$ and obtaining the conventional scale-free continuum theory,  we preserve in the theory a small but finite value of the parameter $\delta$. In this way we effectively account for finite size effects. In particular, the size of the dislocation cores is overestimated while the number of dislocations is underestimated. However, the resulting approach  preserves the basic nature of  both long-range and short-range interactions of dislocations  at least when they are  away from the boundaries. For instance, the  blown up dislocations   will still  nucleate and self-lock adequately. 

In our numerical code we used dimensionless parameters $\tilde h=h/b=1$ and $\tilde L=L/b=1/N$, where the $b$ is the size of the mesoscopic particle. For the choice of the   scale $b$ it is natural to require  that  the  Cauchy-Born   energy density   computed for a lattice fragment with  scale $b$   exhibits a high level of  periodicity within  the  range of strains reached in a particular  numerical experiment.  We have checked that in our case the assumption that $b \sim 10 a$ is sufficient.

 \paragraph{Numerical method.}
 
  Numerical implementation of the MTM approach reduces to solving an elastic finite element problem.  The goal is to follow the displacements of the   network of discrete nodes  labeled by integer-valued coordinates $I =1,..., N^2$. 
  
  Specifically, we assume that each  of the  elements  is a deformable triangle and  employ the standard  linear  3-node elements  \cite{Irons1966,Bramble1970-oq}. The displacement field  inside each of  the elements,  represented  in   dimensionless form,  can be written in the form
%\begin{equation}
 ${\bf u} ({\bf z}) = \sum_{a } \mathcal{N} ^a({\bf z};h) {\bf u}^a,$
 %\end{equation}
 where ${\mathcal N}^a(\bf z;h)$ are   compactly supported  linear shape functions,  ${\bf u}^a$ are the are the
nodal displacements and summation is assumed over repeated indexes; the interpolation functions for each element are defined in terms of a local  dimensionless coordinate system. The mesoscopic deformation gradient is then 
 %\begin{equation}
 ${\bf F}({\bf z}; h) =\mathbb{I}+ {\bf u} ^a\otimes\nabla\mathcal{N} ^a({\bf z};h).$ 
% \end{equation}

 In terms of macroscopic reference coordinates,  the elastic energy inside each element can be  computed using the simplest   quadrature scheme 
%\begin{equation}
$w ({\bf x}; h)=\frac{1}{2}  \phi({\bf F} ({\bf x}; h)) J({\bf x},h),$
%\end{equation}
where $J$ is the Jacobian of the transformation from local ($\bf z$)   to  global ($\bf x$) coordinate system. Note that in view of our assumptions the  energy of a  deformed triangular element depends on  three parameters:  the deformed lengths  of the bonds   and   the deformed value of the angle between them.

Finding  a solution of an elastic problem implies  local minimization of the energy 
%\begin{equation}
$W=\int_{\Omega}w({\bf x}; h)  d\mathbf{x}, $
%\end{equation}
which is prescribed on a triangulated  domain \(\Omega\).   The conditions of mechanical equilibrium take the form 
 $\nabla\cdot\mathbf{P}=0$ where we introduced  the first Piola-Kirchhoff stress tensor 
%\begin{equation}
%$\mathbf{P}= 2\mathbf{F} \mathbf{m} \frac{\partial w }{\partial \tilde{\bf C}}\mathbf{m}^T,$
%\end{equation}
 $$\mathbf{P}= 2\mathbf{F} \mathbf{m}{\bf \Sigma}\mathbf{m}^T,$$ with  $${\bf \Sigma} = \begin{bmatrix}
\frac{\partial w}{\partial \tilde{C}{11}} & \frac{1}{2}\frac{\partial w}{\partial \tilde{C}{12}} \\
\frac{1}{2}\frac{\partial w}{\partial \tilde{C}{12}} & \frac{\partial w}{\partial \tilde{C}{22}}
\end{bmatrix}.$$ In the finite element representation these  equations reduce to 
%\begin{equation}
$$\frac{\partial W}{\partial \mathbf{u}^{a}}=\int_{\Omega} \mathbf{P}(\mathbf{x};h) \nabla \mathcal{N}^{a} (\mathbf{x};h) d\mathbf{x} = 0.$$ 
%\end{equation} 

The ensuing nonlinear equilibrium problem is solved numerically using the L-BFGS algorithm~\cite{Bochkanov2013-lk}, which constructs a positive definite approximation to the Hessian, enabling quasi-Newton steps that progressively reduce the total energy. Iterations continue until the energy change per iteration falls below a prescribed tolerance. At each iteration, starting from an approximate solution ${\bf w}^a$, we compute a displacement correction $\text{d}{\bf w}^a={\bf u}^a-{\bf w}^a$ by solving the linearized equilibrium equations via LU factorization~\cite{Sanderson2016-ht,Fishman2022-nj}:
$$K^{ab}_{ij}dw_j^b+R_i^a =0,$$
where$$K^{ab}_{ij}= A_{ipjq}({\bf F}) \frac{\partial {\mathcal N}^a}{\partial x_p}\frac{\partial {\mathcal N}^b}{\partial x_q} $$ is the global stiffness matrix and $$R^a_i= P_{ip}( {\bf F}) \frac{\partial {\mathcal N}^a}{\partial x_p},$$is the residual force vector 
with summation implied over repeated indices. Note that  we introduced  the tensor of tangential elastic moduli
$$A_{iajb}= \frac{\partial^2\phi_\mathscr{D}{({\bf \tilde C} )}}{\partial F_{ia}\partial F_{jb}}.$$ which  can be used construct the Eulerian acoustic tensor
$${Q}_{ij}=F_{la} F_{mb} A_{iajb} n_l n_m,$$
where $\bf n$ is an arbitrary unit vector. The Legendre-Hadamard  condition  which marks the threshold of elastic instability in continuum theory can be  then written in the form
\begin{equation} \label{detQ}
\det (\boldsymbol{{Q}})=0.
\end{equation}
The corresponding  stability boundary is referred to in the main paper and it is illustrated for the  chosen elastic energy landscape  in Fig.~\ref{01}.

 \paragraph{Implementation of the loading conditions. } To simulate an effectively infinite crystal under controlled  macroscopic affine deformation, we use the method of domain replication. Starting from the computational domain $\Omega_0$ with dimensions $L_x \times L_y$,
 % is replicated in all spatial directions to create a periodic array of domains. 
we generate a periodic lattice of domains using  translation vectors  
 $\mathbf{t}_n = \begin{pmatrix} n_x L_x \\ n_y L_y \end{pmatrix},$ where  $\mathbf{n} = (n_x, n_y) \in \{-1, 0, 1\}^2.$ 
%where $\mathbf{n}$ represents the integer lattice coordinates.
% For the 9-domain implementation, we consider all translation vectors within the immediate neighborhood of the central domain. 
 Let $\bar{\bf F}(\alpha )$ denote the macroscopic deformation gradient tensor at  the parameter value  $\alpha$. Under such affine deformation, the position of   replicated domains relative to the central domain $\Omega_0$ is given by:
 $\mathbf{R}_n = \bar{\bf F}(\alpha ) \cdot \mathbf{t}_n,$ 
where $\mathbf{R}_n$ represents the displacement vector of the origin of the domain associated with $\mathbf{t}_n$. 
Then, for a node located at position $\mathbf{x}_i$ in the central domain $\Omega_0$, its periodic image, associated with the  translation vector $\mathbf{t}_n$, is at:
 $\mathbf{x}_i^{(n)} = \mathbf{x}_i + \mathbf{R}_n = \mathbf{x}_i + \bar{\bf F}(\alpha ) \cdot \mathbf{t}_n.$ 
Such correspondence ensures that the macroscopic deformation gradient $\bar{\bf F}(\alpha )$ is consistently applied across all periodic images of $\Omega_0$. After establishing the positions of all nodes across the nine domains (original domain $\Omega_0$ and its eight periodic images), we perform a Delaunay triangulation on the complete set of particles:
 $\mathcal{P} = \{\mathbf{x}_i : i \in \mathcal{I}_0\} \cup \{\mathbf{x}_i^{(n)} : i \in \mathcal{I}_0, n \neq \mathbf{0}\},$ 
where $\mathcal{I}_0$ denotes the set of particle indices in the original domain and $\mathbf{0} = (0,0)$ corresponds to the null translation vector.  Using the obtained global  triangulation $\mathcal{T}$, we select  the  triangles that have at least one vertex belonging to the original domain $\Omega_0$. 
\begin{figure}[h!]
\centering
\includegraphics[scale=0.15]{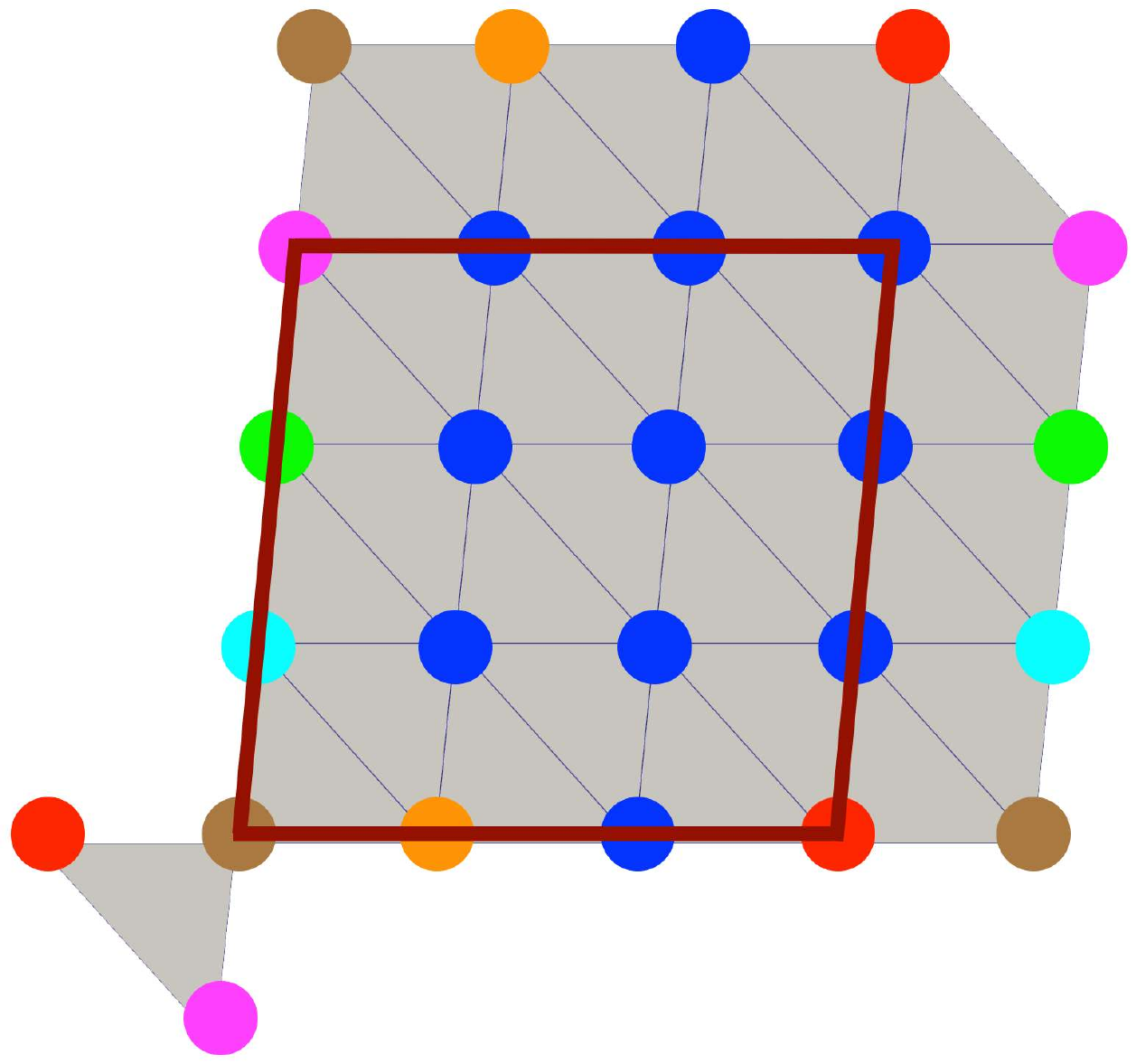}
\caption{ The periodicity unit of the computational mesh $\mathcal{T}_{\text{ch}}$  obtained by  periodic Delaunay triangulation. The central computational domain $\Omega_0$ is limited by the brown box. Only triangles with at least one vertex in $\Omega_0$ are selected for the mesh (shown by the thin lines), ensuring proper connectivity across periodic boundaries while avoiding redundant counting of triangles.
}\label{sm:fig2}
\end{figure}

Note that  since triangles near periodic boundaries may be counted multiple times through different domain images, we need to ensure  uniqueness by mapping all vertex indices to their representatives in $\Omega_0$ and using the sorted tuple of these representatives as a canonical identifier. 

Specifically, a triangle is included in the chosen mesh $\mathcal{T}_{\text{ch}}$ only upon its first encounter, ensuring that each physical triangle is counted exactly once. The ensuing periodicity unit of the computational mesh $\mathcal{T}_{\text{ch}}$  is illustrated in Fig.~\ref{sm:fig2}, where we show the central domain $\Omega_0$ (limited by the brown box) together with   triangles associated with the nodes  constituting this domain.  Such  selection  procedure ensures that:
(i) Each triangle in the computational mesh is counted exactly once, eliminating redundancy from the periodic extension; (ii)
 All triangles connected to the original domain are included, maintaining proper connectivity across periodic boundaries;
(iii) The triangulation naturally incorporates the macroscopic deformation through the deformed positions of the periodic images.

 The selected set of triangles $\mathcal{T}_{\text{ch}}$ was used as the computational mesh in the MTM based numerical experiments  ensuring both, compatibility  with   the periodic boundary conditions and  consistency with the applied macroscopic deformation gradient $\bar{\bf F}(\alpha )$.

\paragraph{Dynamics in the configuration space.}

\begin{figure}[htbp]
\centering
\begin{tabular}{@{}c@{\hspace{2pt}}c@{}}
\begin{tabular}{cc}
\includegraphics[trim=0 0 138 0, clip, scale=0.17]{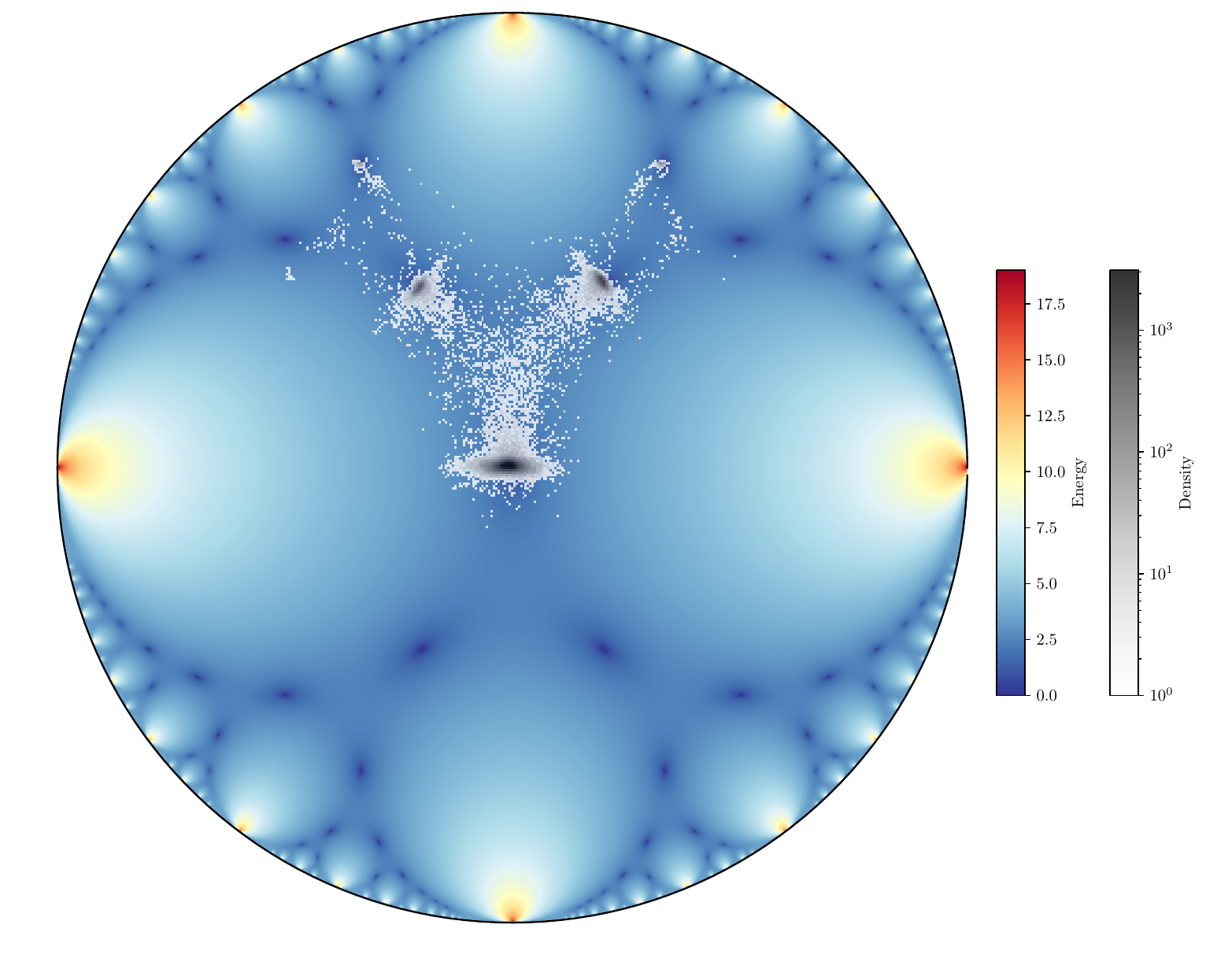} &
\includegraphics[trim=0 0 138 0, clip, scale=0.17]{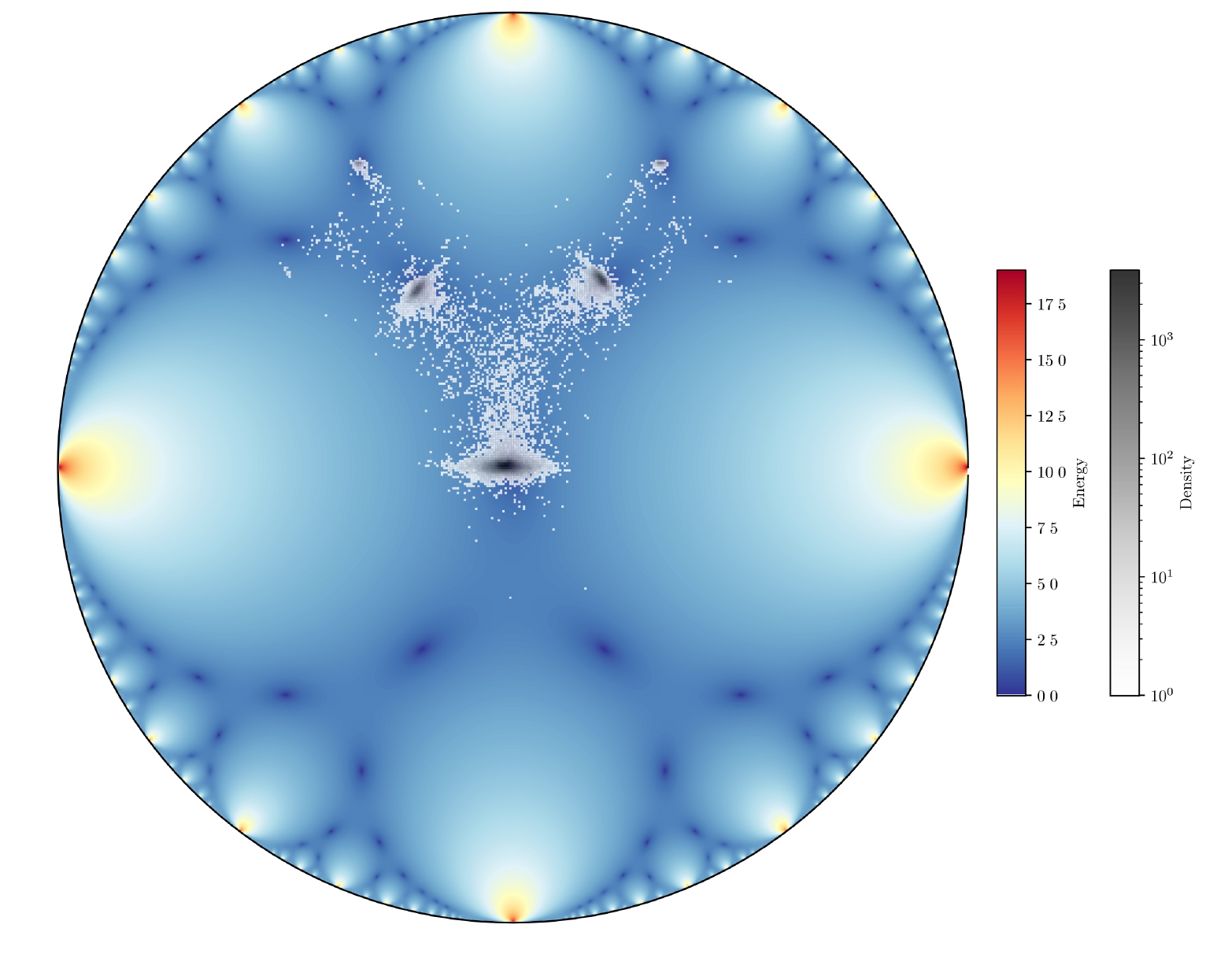} \\
(a) & (b) \\
\includegraphics[trim=0 0 138 0, clip, scale=0.17]{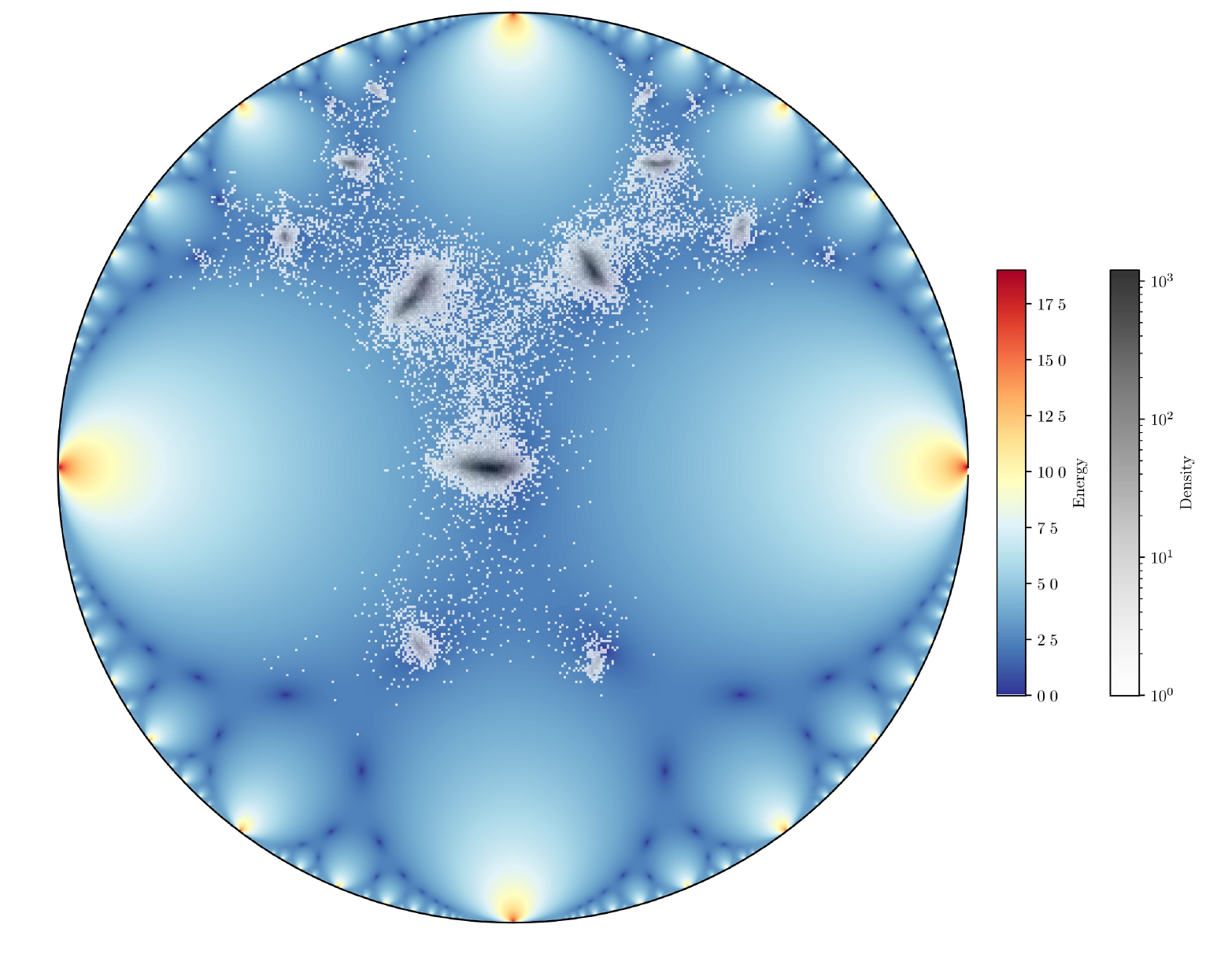} &
\includegraphics[trim=0 0 138 0, clip, scale=0.17]{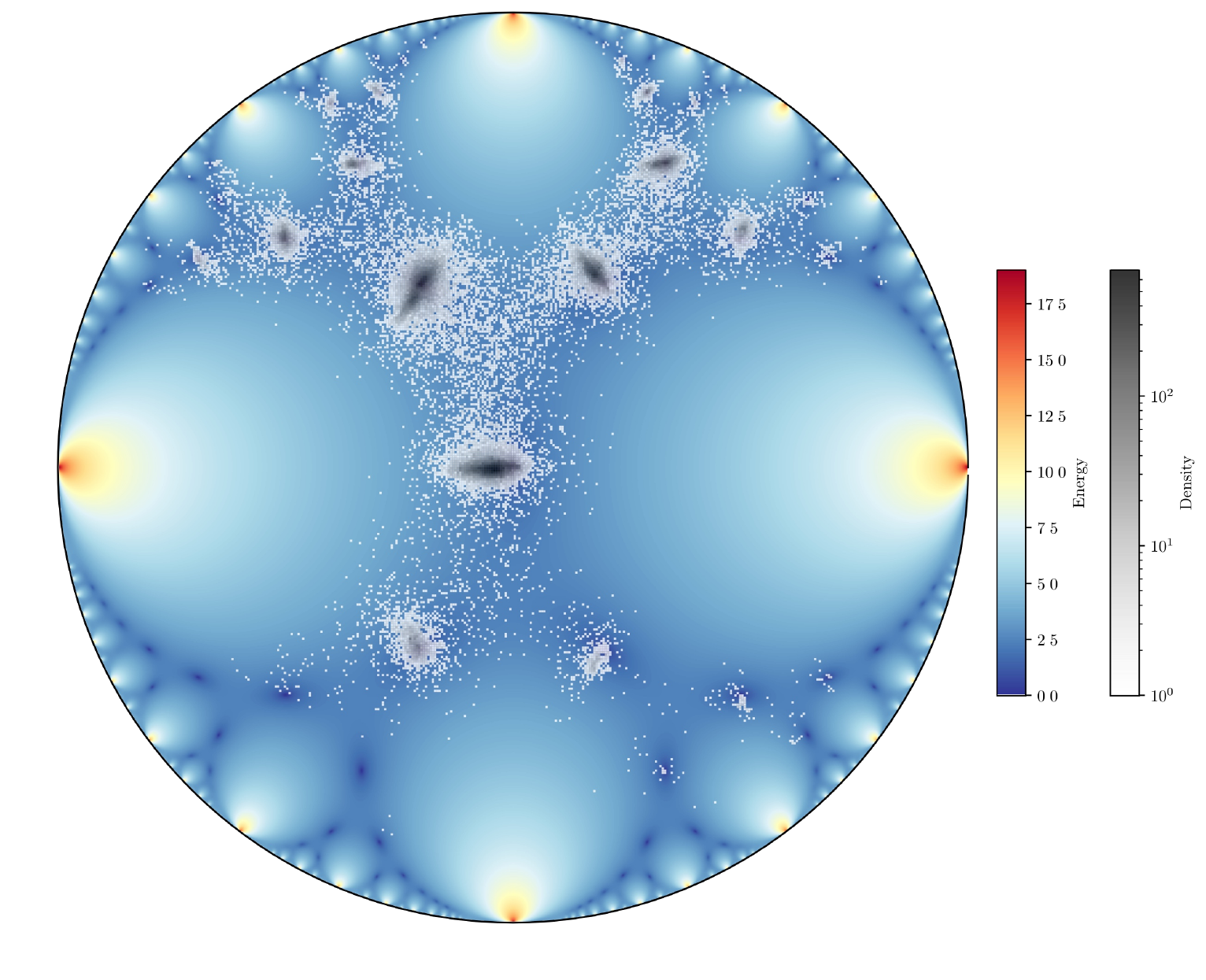} \\
(c) & (d)
\end{tabular} &
\begin{tabular}{c}
\includegraphics[trim=597 0 20 0, clip, scale=0.26]{./Fig3sm-a.pdf}
\end{tabular}
\end{tabular}
\caption{Distribution of values of the local metric tensors in the Poincaré configurational space at four values of loading parameter.  Background color  represents  the energy landscape whose 3D representation was  shown in Fig. 1(b) in the main text.  The color of the gray dots represents the number of the finite elements  sharing  the  same metric tensor. (a) $\alpha=0.3$ (generic pre-yield state), (b) $\alpha=0.5$ (pre-yield state close to the yield threshold), (c) $\alpha=0.6$ (immediate post-yield state), and (d) $\alpha=0.75$ (mature post-yield state). The  states (a,b) show  to the development of the micro-plastic regime and show how small are the associated changes in the configuration of the elements. The (c, d) illustrate the proliferation of plastic strain in the post-yield regime  with the number of  populated energy wells gradually  increasing.}
\label{sm:fig_energy_histogram}
\end{figure}

Here we provide additional illustrations detailing  the evolution of system in the configurational space of local values of the metric tensors projected on Poincaré  disk, see   Fig.~\ref{sm:fig_energy_histogram}  (a-d). In the pre-yield micro-plastic regime, the system predominantly occupies three primary energy wells, corresponding to early-stage, isolated dislocation gliding along the fundamental $x$ and $y$ crystalline orientations. Conversely, in the post-yield regime, accumulated plastic strain and the macroscopic formation of shear bands cause the system to populate a much broader landscape of potential wells. Moreover, the stability envelopes derived from macroscopic acoustic tensor analysis strictly bound the local regions surrounding each distinct well. This clear correlation demonstrates that the vast majority of mesoscopic units remain  localized inside these 'locally' stable  regions, leaving only the highly energetic, actively moving dislocation cores exposed as they traverse the unstable saddles connecting the energy minima.

\paragraph{Finite size scaling and data collapse.}
To determine the power-law exponent and the finite-size scaling exponent (often interpreted as the fractal dimension of the avalanches), we employ a robust data-collapse optimization procedure. 

%Near the critical point, the probability distribution $P(s;L)$ of avalanche sizes $s$, which can represent either the dissipated energy $\Delta W$ or the macroscopic stress drop $\Delta \Sigma $ (defined as the drop in the Cauchy stress tensor component $\sigma_{xy}$ multiplied by the current surface area $A$, yielding $\Delta \Sigma  = A \Delta\sigma_{xy}$), for a system of linear size $L=Nh$ obeys the finite-size scaling (FSS) ansatz:
%\begin{equation}
%  P(s; L) \;=\; s^{-x}\, \mathcal{F}\!\left(\frac{s}{L^{D_x}}\right),
%\end{equation}
%where $x$ represents the relevant power-law exponent ($x = \tau$ for stress drops and $x = \epsilon$ for dissipated energy), $D_x$ is the corresponding fractal dimension ($D_\epsilon$ or $D_\tau$), and $\mathcal{F}(u)$ is a universal scaling function. 

Rather than relying on the algebraic moment spectrum, we directly rescale the axes to collapse distributions from all system sizes onto a single master curve. We define the rescaled variables:
\begin{equation}
  \tilde{x}(s,L) \;=\; \frac{s}{L^{D_x}},
  \qquad
  \tilde{y}(s,L) \;=\; P(s;L)\cdot L^{D_x x}.
\end{equation}
The optimal parameters $(x^*, D_x^*)$ are determined by minimizing the residual spread between the rescaled curves across their common overlapping range. For a discrete set of system sizes $\{L_k\} = N_kh$, the quality of the collapse at a trial pair $(x,D_x)$ is evaluated systematically. First, we compute the discrete logarithmic data points $(u_{k,i}, v_{k,i}) = (\log_{10}\tilde{x}_{k,i}, \log_{10}\tilde{y}_{k,i})$ for each system size $L_k$.

Next, we determine the common logarithmic interval by identifying the overlapping limits $u_{\min} = \max_k[\min_i(u_{k,i})]$ and $u_{\max} = \min_k[\max_i(u_{k,i})]$. For our datasets, a robust overlap domain naturally exists near the optimal scaling exponents. We then define a uniform grid of 80 points $\{c_j\}$ spanning the interval $[u_{\min}, u_{\max}]$. To evaluate the curves on this common grid, we construct a continuous function $\hat{v}_k(u)$ for each system size using linear interpolation between the discrete points $(u_{k,i}, v_{k,i})$.

We define the valid index set $\mathcal{V}$ as the subset of grid indices $j$ such that the grid point $c_j$ lies strictly within the measured domain of all $N_L$ interpolated curves, ensuring no extrapolation is required. The quality of the collapse is quantified by computing the average variance across this valid set:
\begin{equation}
  \mathcal{Q}(x,D_x) 
  = 
  \frac{1}{|\mathcal{V}|} \sum_{j\in\mathcal{V}} \left( \frac{1}{N_L - 1} \sum_{k=1}^{N_L} \left[ \hat{v}_k(c_j) - \bar{v}(c_j) \right]^2 \right),
  \label{eq:quality}
\end{equation}
where $\bar{v}(c_j) = \frac{1}{N_L} \sum_{k=1}^{N_L} \hat{v}_k(c_j)$ is the mean value of the interpolated curves at the fixed grid point $c_j$.

Evaluating this collapse variance provides a highly sensitive measure of how well the distributions coincide. We employ a two-stage grid-search (first coarse, then fine) in the $(x, D_x)$ parameter space to locate the global minimum of the variance, $\mathcal{Q}_{\text{min}}$. The coordinates of this minimum define the optimal scaling exponents, $(x^*, D_x^*)$. 

To estimate the uncertainties of these exponents, we analyze the curvature of the $\mathcal{Q}$ landscape immediately surrounding the minimum. For each parameter, we project the variance onto a 1-D profile and fit it with a parabola. Because traditional statistical error models do not strictly apply to this data-collapse metric, we establish an operational threshold for the error: the $1$-$\sigma$ confidence interval is defined as the parameter deviation required to double the variance from its optimal value (i.e., the half-width where the fit reaches $2\mathcal{Q}_{\text{min}}$). Consequently, a sharp, narrow minimum in the variance landscape clearly demonstrates that the scaling parameters are tightly constrained by the overlapping data.

\begin{figure}[hbt!]
\centering
\begin{tabular}{cc}
(a) & (b) \\
\includegraphics[width=0.21\textwidth]{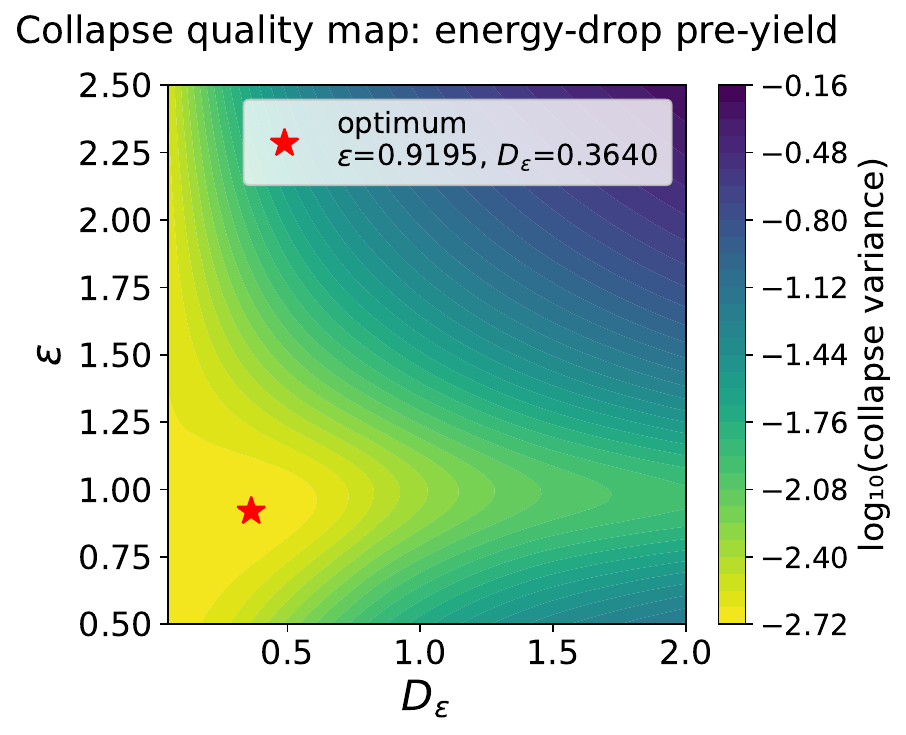} &
\includegraphics[width=0.21\textwidth]{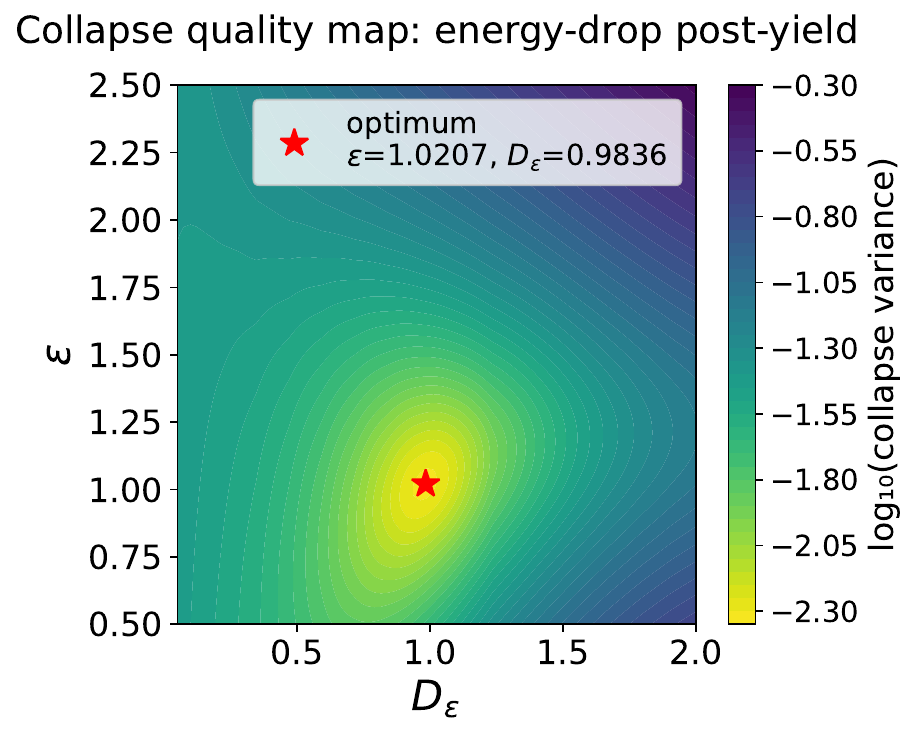} \\
\end{tabular}
\caption{Landscapes  of the quality metric $\mathcal{Q}$ for the pre- and post-yield regimes.  (a)  configuration of  $\mathcal{Q}(\epsilon,D_\epsilon)$ in pre-yield regime, and (b)  similar graph for the post-yield regime. The deep, well-defined global minima in these variance surfaces allow us to choose  the optimal scaling exponents and determine the associated geometric 1-$\sigma$ uncertainties.}
\label{sm:fig_quality_maps}
\end{figure}

\begin{figure}[h!]
\centering
%\begin{tabular}{cc}
%(a) Pre-yield avalanche & (b) Post-yield avalanche \\
%\includegraphics[scale=0.12]{./Fig5-a.pdf} &
\includegraphics[scale=0.18]{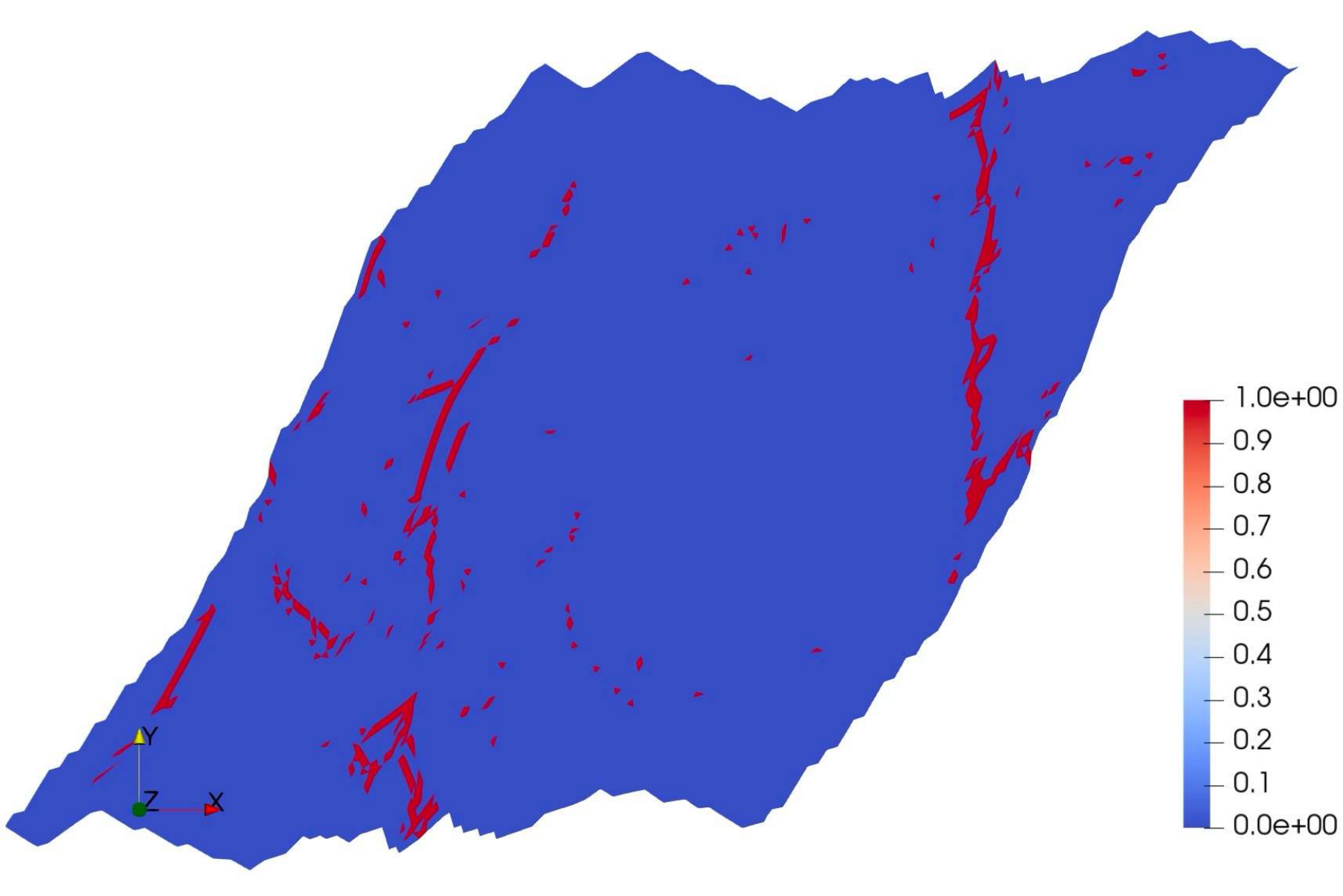} 
%\\
%\end{tabular}
\caption{Visualization of the  geometric morphology of plastically active regions during   typical avalanche events in the post-yield regime. Domains undergoing almost  100$\%$   plastic deformation  are seen as  red.  Plastic deformation is seen to be localized into highly correlated,   system-spanning  branched 1D structures,  with the typical spread  comparable to the linear size of the system. See also the attached supplemental movie 1.}
\label{sm:fig_plastified}
\end{figure}

%\begin{figure}[hbt!]
%\centering
%\vspace{3mm}
%\begin{tabular}{cc}
%\includegraphics[width=0.2\textwidth]{./fig_sm/Fig4-a.pdf} &
%\includegraphics[width=0.2\textwidth]{./fig_sm/Fig4-b.pdf} \\
%(a) Pre-yield (Stress) & (b) Post-yield (Stress) \\
%\includegraphics[width=0.2\textwidth]{./fig_sm/Fig4-c.pdf} &
%\includegraphics[width=0.2\textwidth]{./fig_sm/Fig4-d.pdf} \\
%(c) Pre-yield (Energy) & (d) Post-yield (Energy) \\
%\end{tabular}
%\caption{Optimal data collapses of the probability distributions onto universal master curves. (a, b) Collapse of the macroscopic stress drops $S$ in the pre-yield and post-yield regimes, utilizing exponents $\tau$ and $D_\tau$. (c, d) Collapse of the dissipated energy $\Delta W$ in the pre-yield and post-yield regimes, utilizing exponents $\epsilon$ and $D_\epsilon$.}
%\label{sm:fig_collapses}
%\end{figure}
The variance spread landscapes and the resulting optimal data collapses are presented in Fig.~\ref{sm:fig_quality_maps} and Fig.~6  in the main paper. As can be seen directly from these plots  the scaled probability distributions for different system sizes $L=Nh$ fall remarkably well onto a single universal master curve in both the pre-yield and post-yield regimes. This excellent visual overlap, spanning several decades in both probability and avalanche size, serves as strong empirical validation of the finite-size scaling ansatz and confirms the reliability of the extracted exponents.

\paragraph{Spatial patterns.}

The emerging values of the fractal dimensions of $D_\epsilon$ and  $D_\tau $ in pre and post yield regimes  reflect the difference in spatial structures of the corresponding avalanches. Thus, in the pre-yield microplasticity regime we see individual avalanches represented by   highly localized, spatially scattered, microscopically isolated rearrangements.   Instead, in the  post-yield regime such  seemingly  non connected  plastification zones are replaced by  highly correlated, system-spanning macroscopic structures, with  plasticification zones showing noticeable  branching, see Fig.~\ref{sm:fig_plastified}. 
The  typical avalanches in this regime involve broadening of the shear bands and may include correlated  restructuring of the grain-boundaries whose total length  is comparable with  linear size of the system. One can then talk about the transition in structural topology from almost 0D avalanches (isolated clusters) almost to 1D avalanches (extended anisotropic bands). 
 The suggested  qualitative picture  is, of course,  in full agreement  with the values of the fractal dimension extracted from the data-collapse optimization for pre and post yield regimes.